\documentclass[fleqn,usenatbib]{mnras}

\usepackage[T1]{fontenc}
\usepackage{ae,aecompl}
\usepackage{soul}

\usepackage{newtxtext,newtxmath}
\usepackage{graphicx}	% Including figure files
\usepackage{amsmath}	% Advanced maths commands
\usepackage{amssymb}	% Extra maths symbols
\usepackage[usenames]{color}
\usepackage{natbib}
\usepackage{float}
\DeclareMathAlphabet{\mathpzc}{OT1}{pzc}{m}{it}\definecolor{purple}{RGB}{160,32,240}

\newcommand{\Mh}{M_h}
\newcommand{\Mbh}{M_\bullet}
\newcommand{\Msun}{M_\odot}

\newcommand{\mach}{\mathpzc{M}}
\newcommand{\Vesc}{V_\mathrm{esc}}

\newcommand{\Rvir}{R_{\mathrm{vir}}}

\title{Recoiling Supermassive Black Hole Escape Velocities from Dark Matter Halos}

\author[Choksi, et al.]{Nick Choksi,$^{1}$\thanks{E-mail: nchoksi@berkeley.edu}
Peter Behroozi,$^{1,2}\thanks{E-mail: behroozi@berkeley.edu}$ 
Marta Volonteri$^{3}$,
Raffaella Schneider$^{4}$, 
Chung-Pei Ma$^{1}$, \and
Joseph Silk$^{3,5,6}$, 
Benjamin Moster$^{7}$
\\
$^{1}$Department of Astronomy, University of California at Berkeley, Berkeley, CA, 94720, USA\\
$^{2}$Hubble Fellow\\
$^{3}$Institut d'Astrophysique de Paris, UMR 7095 CNRS, Universit\'{e}
Pierre et Marie Curie, 98bis Blvd Arago, 75014 Paris,
France\\
$^{4}$INAF/Osservatorio Astronomico di Roma, Via di Frascati 33, 00090 Roma, Italy\\
$^{5}$Department of Physics and Astronomy, The Johns Hopkins
University Homewood Campus, Baltimore, MD 21218, USA\\
$^{6}$Beecroft Institute for Cosmology and Particle Astrophysics, University of Oxford, Keble Road, Oxford OX1 3RH\\
$^{7}$Universit{\"a}ts-Sternwarte, Ludwig-Maximilians-Universit{\"a}t M{\"u}nchen, Scheinerstr. 1, 81679 M{\"u}nchen, Germany
}

\date{Released \today}

\pubyear{2017}

\begin{document}
\label{firstpage}
\pagerange{\pageref{firstpage}--\pageref{lastpage}}
\maketitle

\begin{abstract}
We simulate recoiling black hole trajectories from $z=20$ to $z=0$ in dark matter halos, quantifying how parameter choices affect escape velocities. These choices include the strength of dynamical friction, the presence of stars and gas, the accelerating expansion of the universe (Hubble acceleration), host halo accretion and motion, and seed black hole mass.  $\Lambda$CDM halo accretion increases escape velocities by up to 0.6 dex and significantly shortens return timescales compared to non-accreting cases. Other parameters change orbit damping rates but have subdominant effects on escape velocities; dynamical friction is weak at halo escape velocities, even for extreme parameter values. We present formulae for black hole escape velocities as a function of host halo mass and redshift. Finally, we discuss how these findings affect black hole mass assembly as well as minimum stellar and halo masses necessary to retain supermassive black holes.
\end{abstract}

%%%%%%%%%%%%%%%%%%%%%%%%%%%%%%%%%%%%%%%%%%%%%%%%%%
%%%%%%%%%%%%%%%%% BODY OF PAPER %%%%%%%%%%%%%%%%%%
\section{Introduction}
\label{sec:Intro}
At $z=0$, supermassive black holes (SMBHs) have been found in nearly every galaxy, with masses ranging from $10^4$ $\Msun$ to $10^{10}$ $\Msun$ (\citealt{Kormendy2013}; \citealt{Miller2013}). SMBH masses correlate with host galaxy properties such as velocity dispersion, luminosity, and bulge mass (\citealt{HaringRix2004}; \citealt{HK11}; \citealt{mc2013}). Additionally, as far as $z=6$ there is strong similarity between both the cosmic star formation rate and total AGN luminosity and the black hole and stellar mass densities (\citealt{MD2014}; \citealt{Schindler2016}). Many authors interpret these correlations as evidence for the coevolution of SMBHs and their hosts.  

Bright quasars  with inferred SMBH masses of $\sim$$10^{9}-10^{10} \Msun$ have been detected as early as $z\approx 7$ in the Sloan Digital Sky Survey (SDSS; \citealt{Fan2001}; \citealt{Jiang2007}; \citealt{DeRosa2011}; \citealt{Mortlock2011}; \citealt{Wu2015}; \citealt{Wang2015}). These observations suggest that SMBHs can rapidly build up their masses in the $\approx$1 Gyr between the formation of the first stars and galaxies near $z\sim$30 and $z\approx 7$. 

How SMBHs formed and quickly assembled their masses remains an open question. Current hypotheses for SMBH formation suggest that they begin as seed black holes through one of two mechanisms: as the $\sim$$10^2 \Msun$ remnant of a Population III star, or as the $\sim$$10^5M_{\odot}$ result of the direct collapse of a cloud of pristine gas (\citealt{Heger2003}; \citealt{Volonteri_mbh_review}). The latter route is possible only through the collapse of gas clouds that cannot cool by metal-line emission or molecular hydrogen and are therefore unable to fragment (\citealt{Omukai2007}). Such conditions might arise in the early universe ($z\gtrsim10$) when metals are still scarce and a UV ionizing background is present to dissociate any $H_2$ molecules (\citealt{Bromm2003}; \citealt{Shang2009}; \citealt{Valiante2016}). These seed black holes then accrete mass at near or super-Eddington rates at early times (\citealt{VR2005}; \citealt{Ohsuga2007}; \citealt{Madau2014}; \citealt{volonteri_superedd}; \citealt{Pezzulli2016}). SMBHs also grow through repeated mergers with other SMBHs following the mergers of two host galaxies, provided sufficient energy can be dissipated to coalesce the two black holes (see \citealt{Milosavljevic2003}; for futher review of primordial SMBH growth see: \citealt{Latif2016}; \citealt{Johnson2016}; \citealt{Valiante2017}).

However, during the inspiral of two merging SMBHs, the binary system will emit gravitational waves anisotropically due to asymmetries in the merging black holes' masses and spin orientations \citep{Hughes2004}. The merger product receives a kick in one direction to conserve net linear momentum. The magnitude of these kicks increases with the mass ratio between the two SMBHs, with kick velocities as high as 3000 km s$^{-1}$ (\citealt{Campanelli2007}; \citealt{Baker2008}; see Fig. 1 of \cite{Volonteri2010} for a comparison). Within the shallow potentials of small, high-redshift halos, many equal mass mergers will lead to large kicks that will displace or eject the SMBH from their host's center.

\begin{figure}
\vspace{-2.5 mm}
\includegraphics[width=\columnwidth]{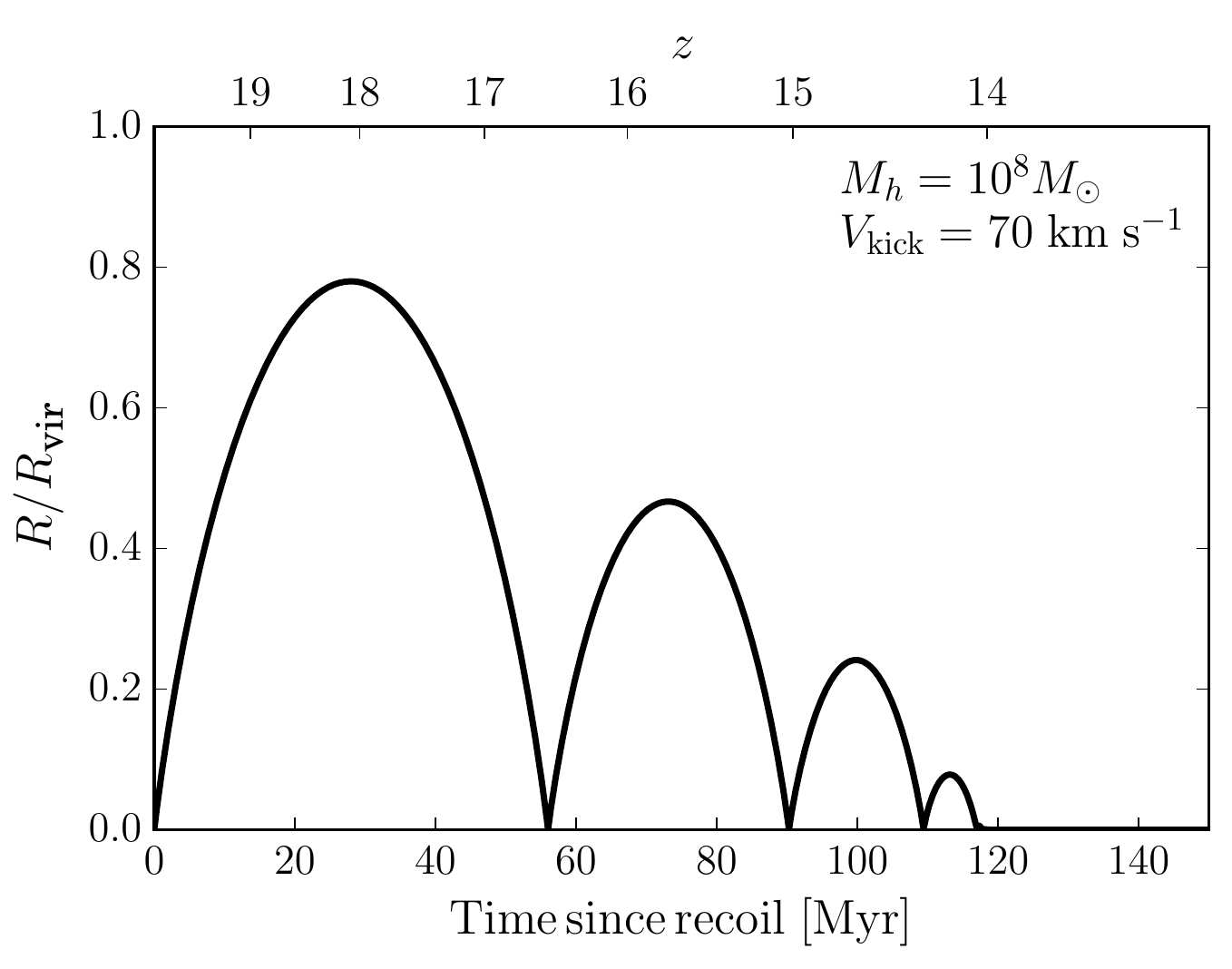}  
\vspace{-4 mm}
\caption{Trajectory of a recoiling supermassive black hole (SMBH), similar to those studied in previous analytic works \citep[e.g.,][]{Madau2004,Volonteri2008,Tanaka2009}. The SMBH oscillates inside the host potential until dynamical friction dissipates its energy. The kick is imparted at $z=20$ to a $10^5 \Msun$ SMBH.}
\label{fig:original}
\end{figure}

Several authors have modeled the trajectories of kicked SMBHs inside static, analytical host dark matter halos (e.g., \citealt{Madau2004}; \citealt{Volonteri2008}; \citealt{Tanaka2009}, hereafter TH09; see Fig.\ \ref{fig:original} for a typical trajectory considered in these works). In particular, previous studies have focused on the possibility that recoil effects could impede the formation of the $\sim$$10^{10} M_{\odot}$ objects suggested by observations and enhance the prospect off-nuclear AGN detection. 

Other groups have used 3D hydrodynamical simulations to follow recoiling SMBHs (e.g., \citealt{Sijacki2011}; \citealt{Blecha2015}), yet the limited resolution of these simulations made modeling dynamical friction effects difficult. \cite{Tremmel2015} introduced a more accurate estimate for unresolved dynamical friction effects. However, they did not consider recoiling SMBHs, which are the main focus of this work.

In this paper, we adopt a hybrid approach. Rather than model halos derived from either semi-analytic merger trees or cosmological simulations, we employ analytical formulae for halo density profiles and dynamical friction, but also include effects previously captured mostly in hydrodynamic simulations -- i.e., host halo accretion and motion as well as Hubble acceleration. A preliminary step in this direction occured with \cite{smole2015}, who considered the effect of analytical accretion rates for a restricted range of halos.  Here, we explore a large parameter space of halo properties and quantify how these affect SMBH escape velocities, and similarly consider effects of varying dynamical friction strength and seed black hole masses.

Our method for following recoiling SMBHs is described in \textsection\ref{sec:Methodology}. Results for how parameter choices affect SMBH trajectories, including fits for escape velocities as a function of host mass and redshift, are in \textsection\ref{sec:Results}. Lastly, \textsection\ref{sec:Conclusions} summarizes and discusses key results.

Throughout, we use $\Mh$ to denote the SMBH's host halo mass (using the virial overdensity definition of \citealt{BN1998}), and $\Mbh$ to refer to the mass of the black hole. We adopt a flat $\Lambda$CDM cosmology, with $\Omega_M = 0.309$, $h = 0.678$, $n_s = 0.968$, $\sigma_8 = 0.816$, $f_b = 0.158$ \citep{Planck2015}.

\section{Methodology}
\label{sec:Methodology}
\subsection{Host halo potential}
\label{subsec:gravity}
\begin{figure}
    \includegraphics[width =\columnwidth]{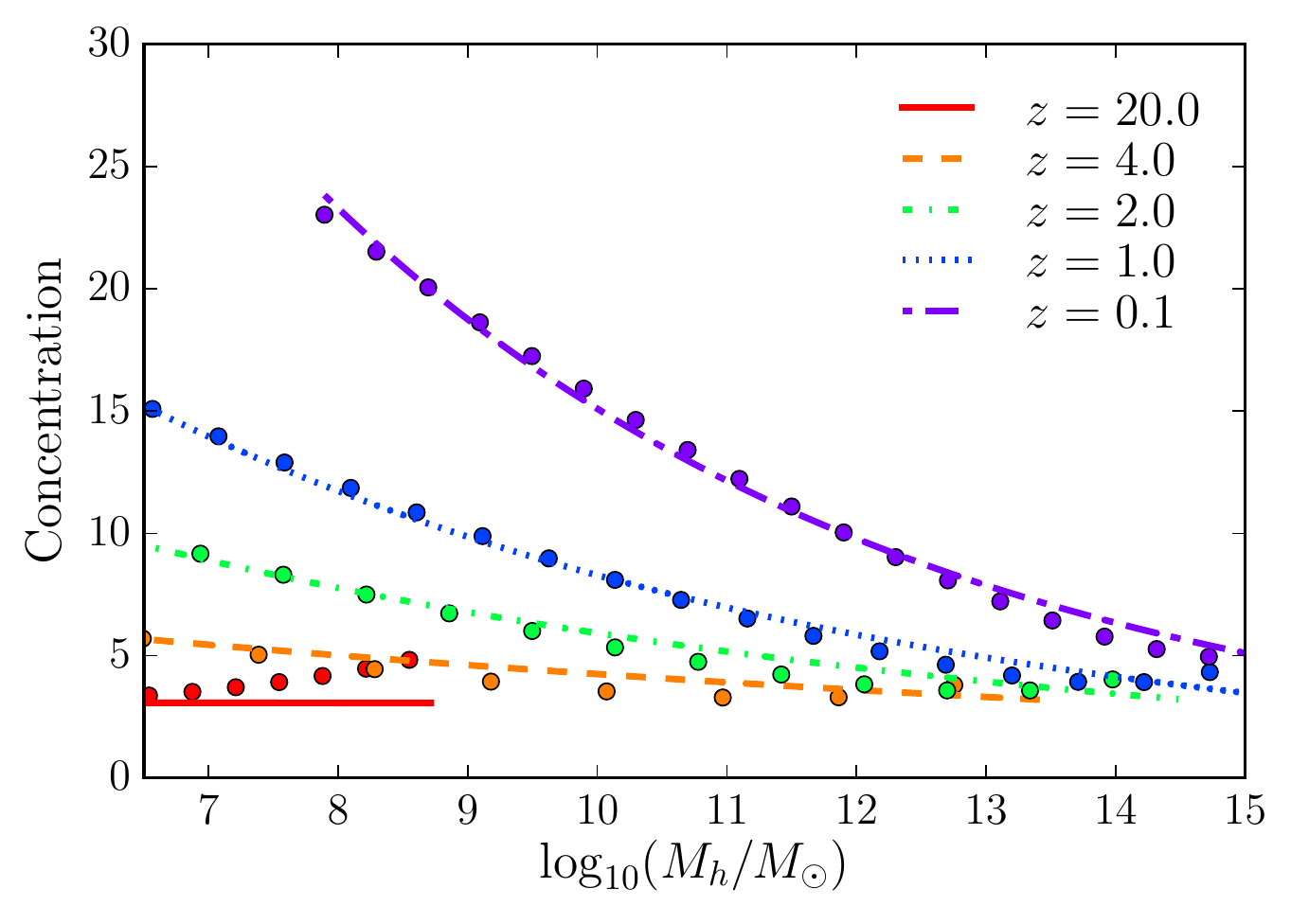}  
    \vspace{-4 mm}
    \caption{Power-law fits (Eqs. \ref{eqn:c1} - \ref{eqn:c3}) to halo concentrations from \protect\cite{Diemer2015}.}
    \label{fig:concentration}
\end{figure}

We model the SMBH's host as a spherically symmetric potential composed of a dark matter halo and superimposed baryonic profile. The dark matter is distributed in a pure NFW profile \citep{Navarro1997}. The variation in halo concentration with host halo mass is well-described by a power law:
\begin{equation}
c(\Mh, z) = c_0(z)\left(\frac{\Mh}{10^{13} \Msun}\right)^{\alpha(z)}. \\
\label{eqn:c1}
\end{equation}
Using results from \cite{Diemer2015}, we fit $c_0(z)$ and $\alpha(z)$ as: 
\begin{eqnarray}
c_0(z) & = & \frac{4.58}{2}\left[\left(\frac{1+z}{2.24}\right)^{.107} + \left(\frac{1+z}  {2.24}\right)^{-1.29}\right] \\
\alpha(z) & = & -0.0965\,\exp\left(-\frac{z}{4.06}\right),
\label{eqn:c3}
\end{eqnarray}
upon which we impose a minimum value of $c=3$. Comparison of our fit with direct results from \cite{Diemer2015} are shown in Fig. \ref{fig:concentration}.

\begin{figure}
\vspace{-1.5 mm}
\includegraphics[width=\columnwidth]{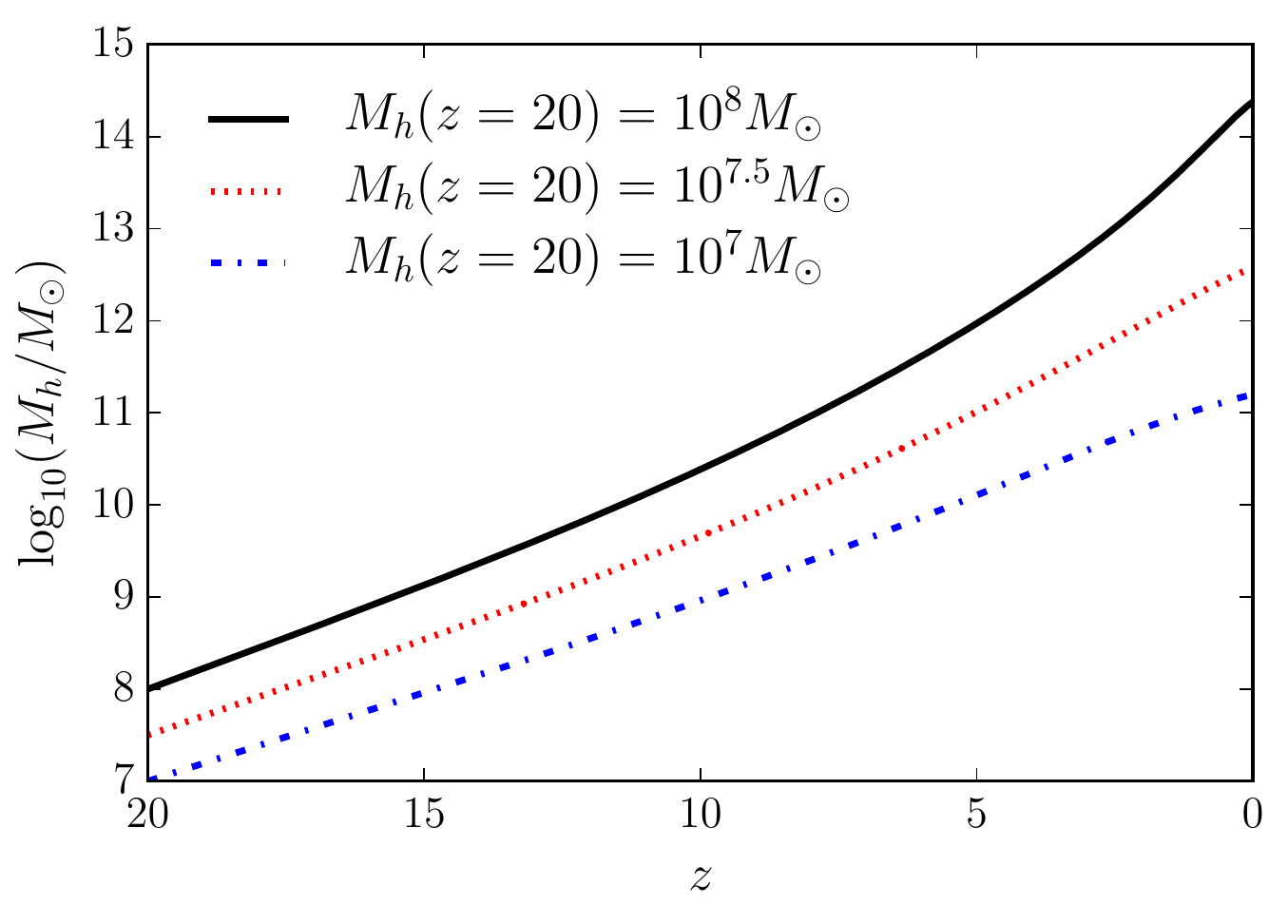}
\vspace{-6.0 mm}
\caption{Mass assembly histories of $10^{7}$ -- $10^8 \Msun$ halos starting at $z=20$, using accretion rates from \protect\cite{Behroozi2015}.}
\label{fig:halogrowth} 
\end{figure}

The host will also grow, both through mergers with other halos as well as through smooth accretion from the intergalactic medium (IGM). As discussed in \S\ref{sec:Intro}, halo growth has previously been modeled only in hydrodynamic simulations and \cite{smole2015}.  An alternative approach is to instead apply an average mass accretion rate, $\dot{\Mh}(\Mh, z)$, for which \cite{Behroozi2015} derived a fitting function using halos in the \textit{Bolshoi} and \textit{Bolshoi-Planck} simulations (Fig. \ref{fig:halogrowth}; \citealt{Klypin2010}; \citealt{Klypin2016}). We examine recoils in a range of halo masses that will grow into $10^{10} \Msun$ to $10^{15} \Msun$ halos at $z=0$, beginning at $z=20$ when the first SMBH seeds are expected to form.

We consider two types of baryon distributions. The total baryonic mass in the host is given by $f_b\Mh$. We begin by assuming all of the host's baryonic mass is gaseous, following an $r^{-2.2}$ density profile with a 1 pc central core of constant density. At high redshifts this is a reasonable assumption, as very few stars have formed. However, we also consider a more realistic model in which the host's stellar mass is set through the stellar mass-halo mass relation given by \cite{Behroozi_sfh}, in a \cite{hernquist} profile with half-mass radius $R_{1/2} = .01\Rvir$ (\citealt{somerville_etal_2014}). Any remaining baryons are then added to the galaxy's gas mass. We discuss the effects of alternate density profiles (both DM and baryonic) in Appendix  \ref{subsec:alternates}.

\subsection{Equation of Motion}
\label{sec:EOM} 
A recoiling SMBH oscillates within its host's potential, governed by the equation of motion: 
\begin{equation} 
\label{eqn:eom}
\ddot{x} = \left(-\frac{G\Mh(x)}{x^2} + a_\mathrm{DF}  - \dot{x}\frac{\dot{\Mbh}}{\Mbh} - qH^2x\right)\hat{\mathbf{x}}.
\end{equation}
The first two terms on the right hand side are the accelerations due to gravity and dynamical friction respectively. The third term gives the change in velocity due to accretion onto the black hole, which causes a decrease in speed to conserve linear momentum. The final term gives the Hubble acceleration, where $x$ is the position of the black hole, and the origin is taken at the center of the host halo.

At the initial recoil redshift, we give the SMBH a radial kick outward from the center of the host ($x=0$). We then numerically integrate Eq. \ref{eqn:eom} using leapfrog integration and a time step of 1000 years. We examine recoiling SMBH behaviour for a range of kicks, up to the escape velocity of the halo.

\subsection{Dynamical Friction}
\label{ss:df}
As a kicked SMBH travels through its host halo, it experiences a drag force due to dynamical friction (DF). The DF acceleration has contributions from both collisionless (dark matter and stars) and collisional (gaseous) material  in the surrounding medium. For an SMBH moving at speed $v$, the collisionless component is given by the standard Chandrasekhar formula \citep{Binney1987}:
\begin{eqnarray}
a_\mathrm{DF}^\mathrm{DM} & = & -\frac{4\pi G^2}{v^2} \Mbh \rho(x) \times \ln\Lambda \left( \mathrm{erf}(\mathrm{X}) - \frac{2}{\sqrt{\pi}}\mathrm{X}e^{-\mathrm{X}^2} \right)\hat{\mathbf{v}} \label{e:df_dm}\\
X & \equiv & \frac{|v|}{\sqrt{2}\sigma_{DM}}.
\label{eqn:df_dm}
\end{eqnarray}
$\rho(x)$ is the sum of dark matter and stellar densities at the BH's position. $\sigma_{DM}$ is the local dark matter velocity dispersion, but varies little over the entire radius of the host. We therefore follow TH09 and use a simplified prescription for a singular isothermal sphere (SIS), $\sigma_{DM} = \sqrt{GM/2R_{\mathrm{vir}}}$. The Coulomb logarithm, $\ln\Lambda$, is not precisely known but is generally taken in the range 2-4 (\citealt{Escala2004}; \citealt{Gualandris2008}). We approximate $\ln\Lambda$ as:
\begin{align}
\ln\Lambda \sim \ln\left(\frac{R_{\mathrm{max}}}{R_{\mathrm{min}}}\right) \sim \frac{1}{3}\ln\left(\frac{\Mh}{\Mbh}\right)
\label{eqn:clog}
\end{align}
and fiducially adopt $\ln\Lambda = 2.3$, corresponding to $\Mh = 10^{8} \Msun$ and $\Mbh = 10^{5} \Msun$ in Eq. \ref{eqn:clog}; alternate values ranging from $\ln \Lambda = 2$ to $20$ do not affect the results (\textsection\ref{subsec:results_df}). 

The DF due to the surrounding gaseous medium is more complicated. Because gas is collisional, it can cool and a larger wake can form behind a traveling SMBH, amplifying the standard DF force. \cite{Ostriker1999} derived an analytical formula for this effect. However, this formula overestimates the drag force at subsonic velocities (TH09). \cite{Escala2004} investigated this problem using numerical simulations and fit a modified prescription of the Ostriker formula by taking a variable value of the Coulomb logarithm. Defining $\mach$ as the Mach number ($\equiv|v|/c_s$), they reduce the drag force by a factor of 2 at $\mach<0.8$, but increase the drag by a factor of 1.5 at $\mach > 0.8$. However, the Escala formula overestimates the amplification for highly supersonic speeds (TH09). We therefore follow TH09 and adopt a hybrid prescription, using the Escala prescription for $\mach < \mach_{eq}$ and and the Ostriker prescription for $\mach > \mach_{eq}$, where $\mach_{eq}$ is the Mach number where the two prescriptions predict the same value for the drag force (here, $\approx 1.7$) The resulting DF acceleration is given by: 
\begin{equation}
\label{e:df_gas}
a_\mathrm{DF}^\mathrm{gas} = -\frac{4\pi G^2}{v^2} \Mbh \rho_\mathrm{gas}(r) \times f(\mach)\hat{\textbf{v}},
\end{equation}
with
\begin{equation} 
f(\mach) = \begin{cases} 0.5\ln\Lambda \left[\mathrm{erf}\left(\frac{\mach}{\sqrt{2}}\right) - \sqrt{\frac{2}{\pi}}\mach e^{-\mach^2/2} \right] &\text{if }  \mach \leq 0.8 , \\ 
1.5\ln\Lambda \left[\mathrm{erf}\left(\frac{\mach}{\sqrt{2}}\right) - \sqrt{\frac{2}{\pi}}\mach e^{-\mach^2/2}\right] &\text{if } 0.8 \leq \mach \leq \mach_{eq} , \\
.5\ln(1 - \mach^{-2}) + \ln\Lambda &\text{if } \mach > \mach_{eq}.
\end{cases}
\end{equation}
The value of $f(\mach)$ depends on the local sound speed, $c_s$, which in turn depends on the local temperature. However, $c_s$ is always less than half the SMBH escape velocity, even at high $z$. For  $\mach > 2$, $f(\mach) \approx \ln\Lambda$. Furthermore, numerical simulations show that the temperature inside the halo should vary by at most a factor of $\approx 3$ \citep{Machacek2001}. The sound speed scales as $\sqrt{T}$, so $c_s$ should vary no more than a factor of two over the entire halo. We therefore do not compute an explicit temperature gradient. Instead, we adopt the prescription of TH09 and assume the entire halo to be isothermal at the virial temperature. $c_s(\Mh, z)$ is then:
\begin{equation}
c_s \approx 1.8(1+z)^{1/2}\left(\frac{\Mh}{10^7\Msun}\right)^{1/3}\left(\frac{\Omega_M h^2}{.14}\right)\mathrm{km}\; \mathrm{s}^{-1}.
\end{equation}
\subsection{Accretion onto the SMBH}
\label{subsec:accretion}
As the SMBH accretes from the surrounding medium, its speed decreases to conserve linear momentum. We again follow TH09 and assume the SMBH undergoes Bondi-Hoyle-Littleton (BHL) accretion, given by \cite{Bondi1944} as:
\begin{equation}
\dot{\Mbh}(r, v) = \frac{4\pi G^2 \rho_b(r) \Mbh^2}{(c_s ^2 + v^2)^{3/2}}.
\end{equation}
We cap accretion at the Eddington luminosity, given by: 
\begin{equation}
\dot{\Mbh} = \frac{1-\epsilon}{\epsilon}\frac{\Mbh}{t_\mathrm{Edd}},
\end{equation}
where $\epsilon$ is the radiative efficiency of the SMBH, generally taken as 0.1, and $t_\mathrm{Edd} = 440\, \mathrm{Myr}$ gives the e-folding time for a black hole accreting at the Eddington rate.

BHL accretion overestimates growth at later times due to local gas depletion and self-regulating feedback \citep{SomDave2015}. Additionally, as discussed in $\S$\ref{sec:Intro}, many authors have suggested that super-Eddington accretion rates are possible. However, when a recoiling SMBH is displaced from the center of its host, accretion rates are negligible and only approach the Eddington limit for short periods as the SMBH passes through the center. So, accretion plays only a minor role in shaping recoil trajectories and the BHL formalism is a sufficient approximation prior to return.
\subsection{Cosmological acceleration}
An SMBH displaced from the center of its host halo will have an effective acceleration from cosmological expansion given by $-qH^2x\hat{\textbf{x}}$, where $q\equiv -\frac{\ddot{a}a}{\dot{a}^2}$ \citep{Nandra2012}. Both $q$ and $H$ evolve with time. Prior to $a \sim 0.6$ ($z\sim 0.68$), cosmological expansion decelerates, causing the SMBH to accelerate back towards the center of the halo.  Afterwards, the expansion accelerates due to the increasing fraction of dark energy (see Fig. \ref{fig:qh2} for the evolution of $-qH^2$), at which point the black hole accelerates away from the center of the host. 
\subsection{Dark matter simulations and host halo motions}
To test the effects of host-halo motion, we use halos from \textit{Bolshoi-Planck}, a dark matter-only simulation within a 250 $\mathrm{Mpc}$ $\mathrm{h^{-1}}$ comoving, periodic box \citep{Klypin2016}. Halos in the simulation were identified using the \textsc{rockstar} code and merger trees were constructed using \textsc{consistent trees} (\citealt{rockstar}; \citealt{ctrees}; \citealt{Bolshoi}). \textit{Bolshoi-Planck} adopts $\Omega_{M} = 0.307$, $h = 0.678$,  $n_s = 0.96$, $\sigma_8 = 0.823$, $f_b = 0.156$, very similar to the parameters in our main analysis.

To estimate the magnitude of the effect caused by movement of the host halo, we choose several $z=0$ halos from the simulation. We then use the peculiar velocities of the halos along the main progenitor branch (MPB) of the chosen halo to track the movement of the host. When the SMBH is kicked, its velocity is decoupled from that of its host. 
However, bulk  external accelerations affect both the host and the SMBH. To cancel this motion, at each simulation output we subtract from the host halo's peculiar velocity the mass-weighted average of the peculiar velocities of all other halos in its progenitor history ($v_\mathrm{host} = v_\mathrm{MPB} - \langle v_\mathrm{progenitors}\rangle$) and spline interpolate at intermediate times.

\section{Results}
\label{sec:Results}
We consider the effect of several parameters on recoiling SMBH trajectories and escape velocities. Throughout, we define the escape velocity as the minimum kick required such that the apocenter of the SMBH's orbit remains outside $0.1R_{\mathrm{vir}}$ after a given amount of time. Specifically, we consider the cases where the SMBH has until $z=0$, until $z=6$, and after 10\% the age of the universe at the time of the kick (i.e., until $1.1t_{\mathrm{kick}}$) to satisfy this criterion.

In \textsection \ref{subsec:results_df} and \textsection \ref{subsec:results_qh2}, we examine the relatively minor effects of varying the Coulomb logarithm and including Hubble acceleration. In \textsection \ref{subsec:results_accretion} we add accretion onto the host halo. This quickly damps the orbits of recoiling SMBHs and makes permanent escape far more difficult. \textsection \ref{subsec:results_baryons} examines the effect of including stars in the host halo. \textsection \ref{subsec:results_mbh} considers variations in the mass of the recoiling SMBH. \textsection \ref{subsec:results_hostmove} examines the trajectories of SMBHs kicked from inside cosmological halos. \textsection \ref{subsec:results_zkick} considers varying the redshift at which the kick occurs.  Finally, \textsection \ref{subsec:formulae} provides formulae for escape velocities as a function of host halo mass and redshift. Throughout, we adopt $\ln\Lambda = 2.3$, $\Mbh = 10^5 \Msun$, and assume the kick occurs at $z=20$ unless otherwise specified.

\subsection{Sensitivity to the Coulomb Logarithm}
\label{subsec:results_df}
\begin{figure}
\vspace{-9 mm}
\includegraphics[width=\columnwidth]{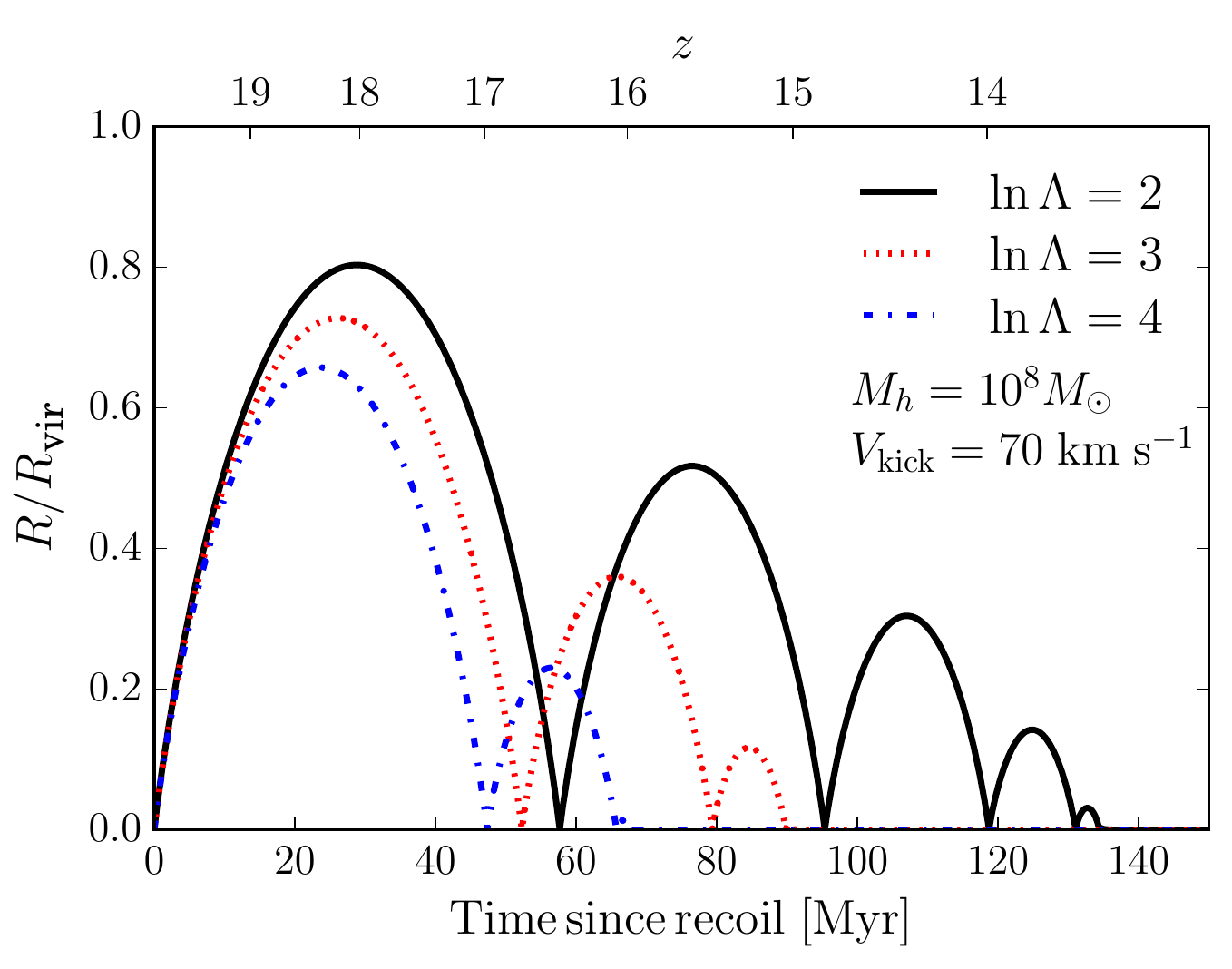}
\vspace{-7.9 mm}
\caption{SMBH trajectories for varying values of the Coulomb logarithm ($\ln\Lambda$; we fiducially adopt $\ln\Lambda = 2.3$). As in Fig. \ref{fig:original}, the kick is at $z=20$ to a $10^5 \Msun$ SMBH.}
\label{fig:df_comparison}
\end{figure}
\begin{figure}
\vspace{-2.5 mm}
\includegraphics[width=\columnwidth]{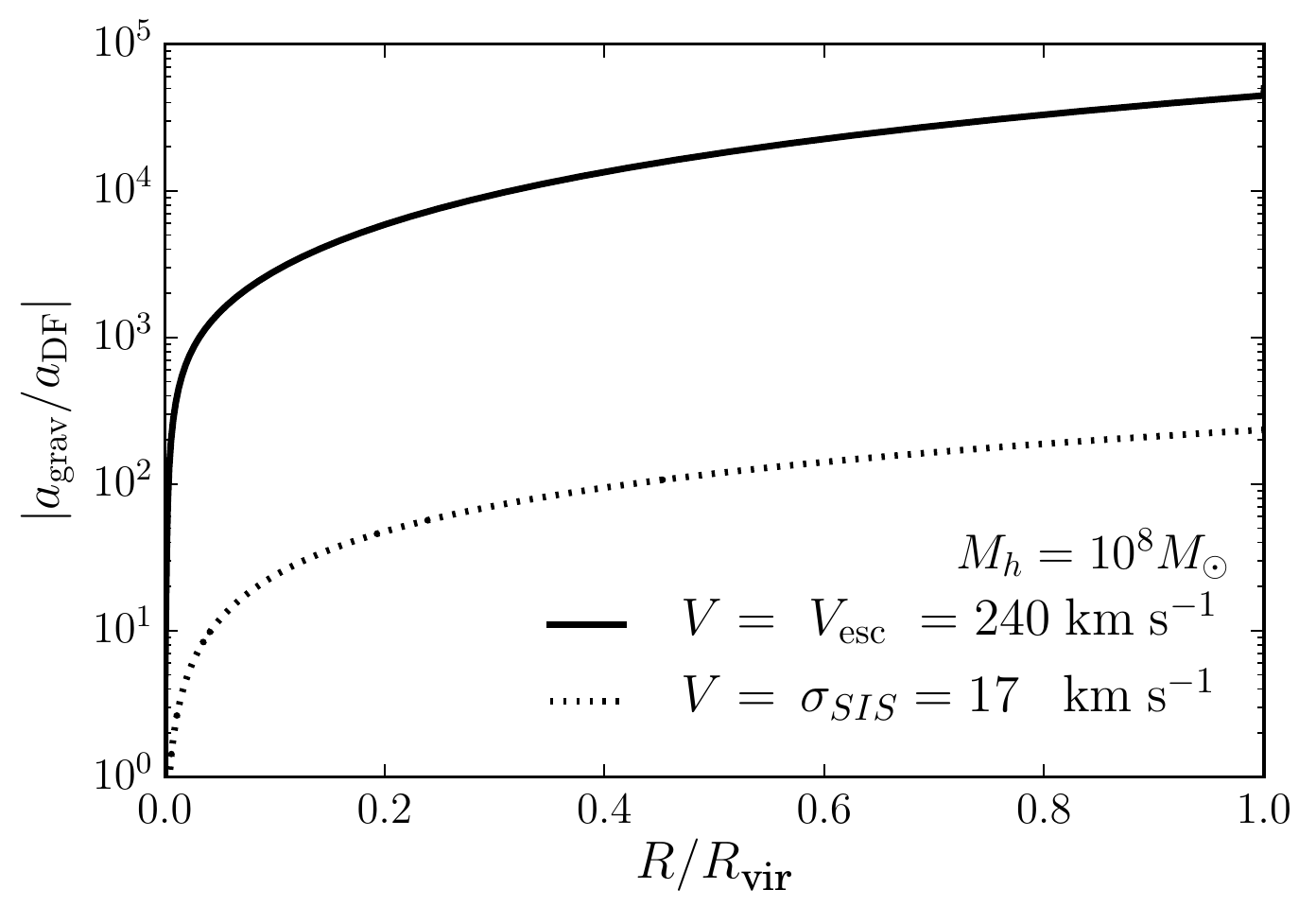}
\vspace{-6.5 mm}
\caption{Dynamical friction is subdominant to gravitational forces, so the host halo's potential well depth largely determines the velocity needed to escape. The figure shows the ratio of gravitational acceleration to dynamical friction acceleration as a function of radius for a $10^5 \Msun$ SMBH traveling at both the escape velocity and velocity dispersion of the halo (solid and dotted lines respectively).}
\label{fig:aratio}
\end{figure}
\begin{figure} 
%\vspace{7.5 mm}
\includegraphics[width=\columnwidth]{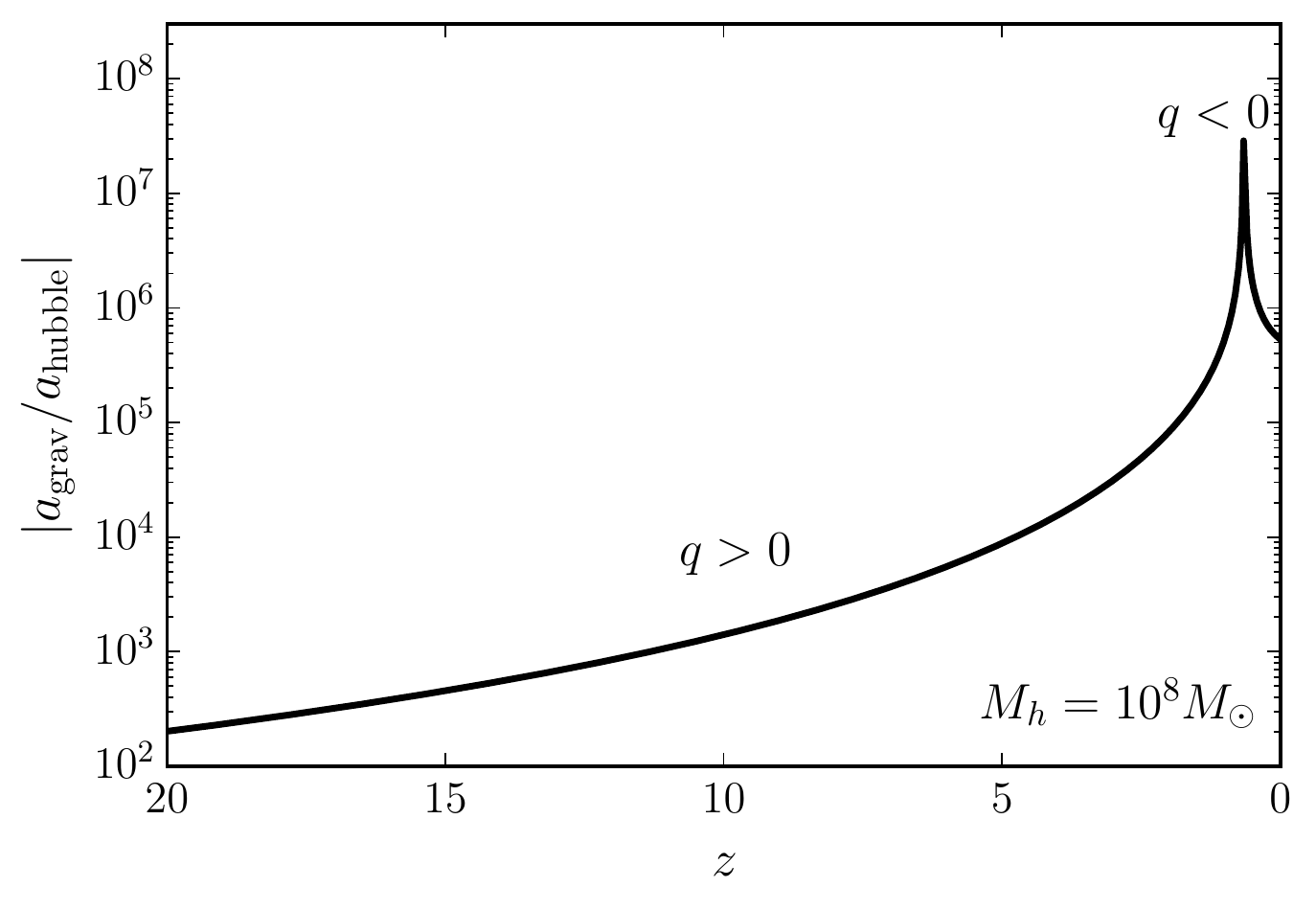}
\vspace{-6.0 mm}
\caption{The Hubble acceleration is negligible at all times. The figure shows the ratio of the gravitational to cosmological acceleration ($qH^2x$), evaluated at the virial radius of a $10^8 \Msun$ halo, as a function of redshift. $q$ changes sign at at z $\approx$ .675. Before this time, the universe is decelerating ($q > 0$) and the Hubble acceleration points towards the center of the halo.} 
\label{fig:qh2}
\end{figure}
\begin{figure}
\vspace{-3 mm}
\includegraphics[width=\columnwidth]{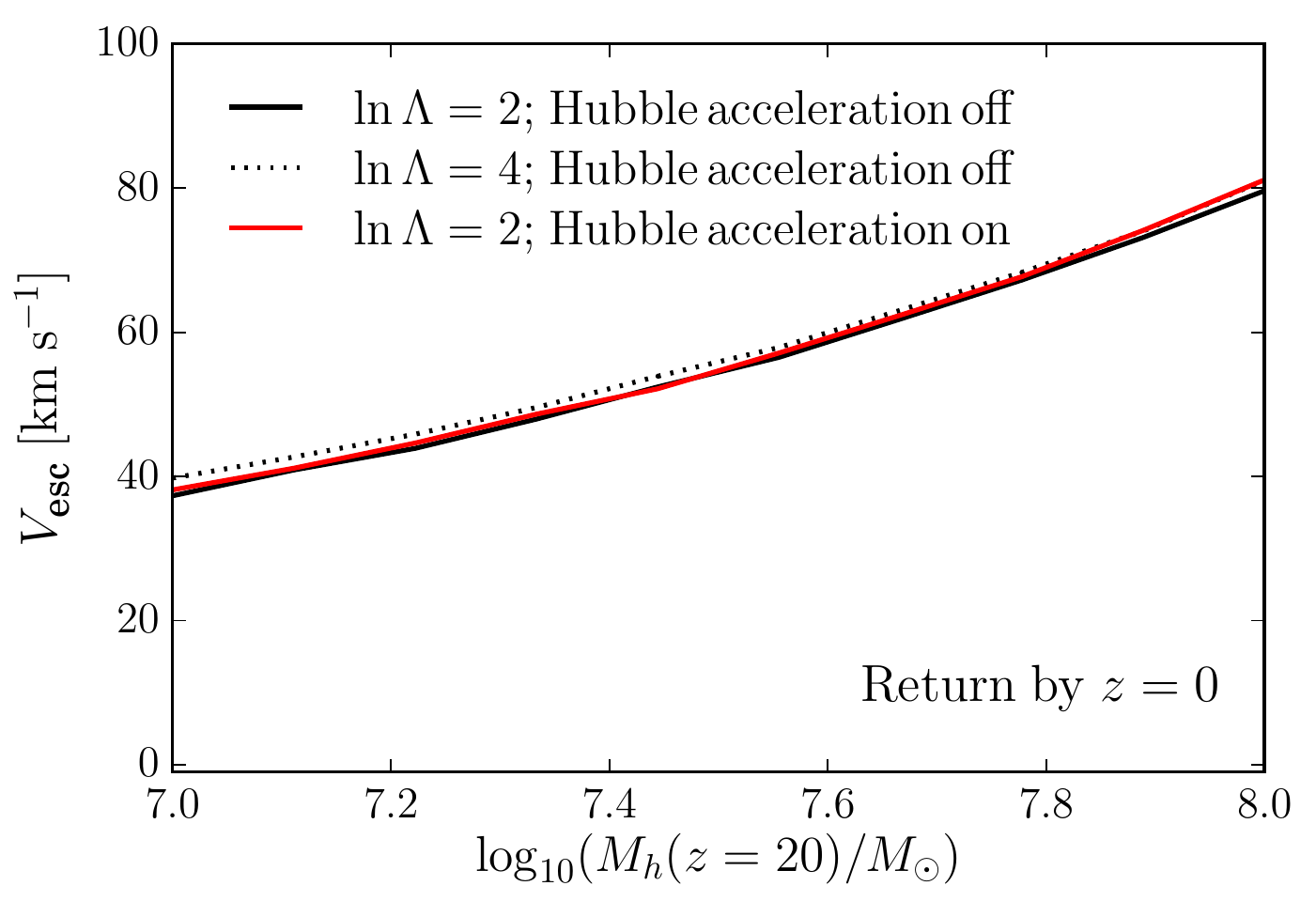}
\vspace{-5.5 mm}
\caption{The velocity needed to escape accreting halos (see return criterion described at the beginning of \textsection\ref{sec:Results}) is negligibly affected by the strength of dynamical friction and the Hubble acceleration. Values of the Coulomb logarithm ranging from 2-4, the extremes of those adopted by previous analytic works, result in $\lesssim$ 0.01 dex change in $\Vesc$ for a kick at $z=20$ to a $10^5 \Msun$ SMBH.}
\label{fig:minor_effects}
\end{figure}
\begin{figure}
\vspace{-6 mm}
\includegraphics[width=\columnwidth]{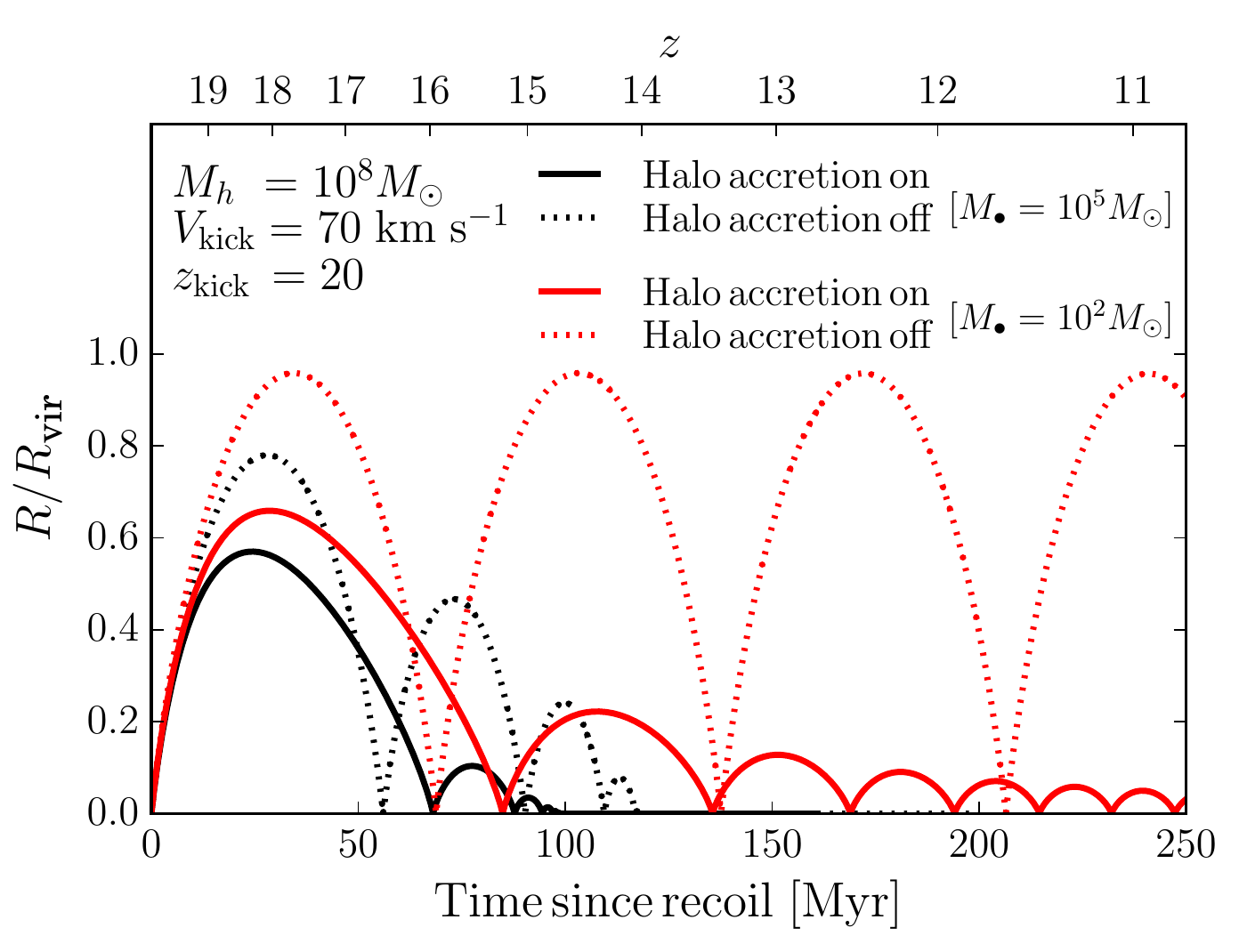}
\vspace{-7 mm}
\caption{SMBH trajectories in an accreting and non-accreting halo for SMBHs kicked at $z=20$. For the $10^2 \Msun$ non-accreting case, DF effects are so small that the BH's orbit shows almost no decay. In the accreting case, the asymmetry of the first peak is due to DF, which peaks for $v \approx \sigma_{DM}$.}
\label{fig:acc_comparison1}
\end{figure}
%halo growth figure was here
The overall shape of SMBH orbits remains unchanged when varying the Coulomb logarithm, with increases in $\ln\Lambda$ mainly leading to faster return to center because dynamical friction (DF) forces are stronger (Fig.\ \ref{fig:df_comparison}). 

At typical halo escape velocities, DF is subdominant to gravitational acceleration (Fig. \ref{fig:aratio}). For $X \equiv v/\sigma \gtrsim 2$ (a reasonable assumption for escape velocities; see Fig. \ref{fig:vesc}), terms with X in Eq. \ref{eqn:df_dm} are negligible:
\begin{equation}
a_{DF}  \approx  \frac{4\pi G^2}{v^2}\Mbh \rho(r) \ln\Lambda.
\end{equation}
Approximating the NFW profile as an isothermal power-law, i.e., 
$\rho(r) = \widetilde{\rho}_0r^{-2}$, we can write the ratio of accelerations as:
\begin{align}
%\left|\frac{a_g}{a_{DF}}\right| & \sim & \left(\frac{\Mh(x)}{\Mbh}\right)\left(\frac{v^2}{Gx^2\rho(x)}\right) \\
\left|\frac{a_g}{a_{DF}}\right|  &\sim  \frac{C_1}{\ln \Lambda}\left(\frac{\Mh(r,t)}{\Mbh(t)}\right)\left(\frac{v}{\sigma}\right)^2\left(\frac{\Mh(r,t)}{r\widetilde{\rho}_0}\right) \label{eqn:ag_adf1} \\
&\sim  \frac{C_2}{\ln \Lambda}\left(\frac{\Mh(r,t)}{\Mbh(t)}\right)\left(\frac{v}{\sigma}\right)^2, \label{eqn:ag_adf2}
\end{align}
where $C_1$ and $C_2$ are constants of order unity and the last line follows because $\Mh(r) \approx \frac{4}{3}\pi\widetilde{\rho}_0r$. The second term in Eq. \ref{eqn:ag_adf2} is $\gg 1$ for all but very small radii or $\Mbh \sim \Mh$. As a result, $|a_g/a_{DF}| \gg 1$, and the escape velocity is not sensitive to the value of $\ln \Lambda$ and for most cases of interest DF gradually damps orbital energy from the SMBH over multiple orbits, as in Fig. \ref{fig:df_comparison}. Instead, the key factors influencing escape velocities are the depth of the halo potential well and its growth over time (Fig. \ref{fig:minor_effects}). 
\subsection{Trajectories in an accelerating universe}
\label{subsec:results_qh2}
Gravitational forces from the dark matter halo always dominate over the Hubble acceleration (Fig. \ref{fig:qh2}), so the latter only marginally affects SMBH orbits. For $q>0$ (decelerating universe), the return-to-center time for most kicks decreases by up to a few percent. The effect's importance increases as $V_{\mathrm{kick}}$ approaches $\Vesc$, as larger kicks can escape to larger radii where the Hubble acceleration is stronger relative to gravity. However, the change in escape velocity is negligible: for kicks at $z=20$, $\Vesc$ increases at most 0.01 dex (Fig. \ref{fig:minor_effects}). 
\subsection{Host halo accretion}
\label{subsec:results_accretion}
Accretion onto the host halo significantly alters the orbits of recoiling SMBHs (Fig. \ref{fig:acc_comparison1}). Accretion onto the host brings the SMBH back to the center faster and makes escape to large radii harder. At high redshift the effect of halo accretion is more pronounced because halos are increasing in mass quickly. For example, in the 100 Myr the $10^5 \Msun$ SMBH in Fig. \ref{fig:acc_comparison1} oscillates inside its host, $\Mh$ increases by $\approx$1 dex (Fig. \ref{fig:halogrowth}). From Fig. \ref{fig:acc_comparison1} it is also clear that host halo accretion is especially important for low mass SMBH seeds because DF is not effective at dissipating energy in this regime. Finally, the effect of host accretion becomes more noticeable as $V_{\mathrm{kick}}$ approaches the escape velocity, because larger kicks have more energy and take longer to have their energy dissipated, allowing more time for the halo to grow.

\begin{figure}
\vspace{0.5 mm}
\includegraphics[width=\columnwidth]{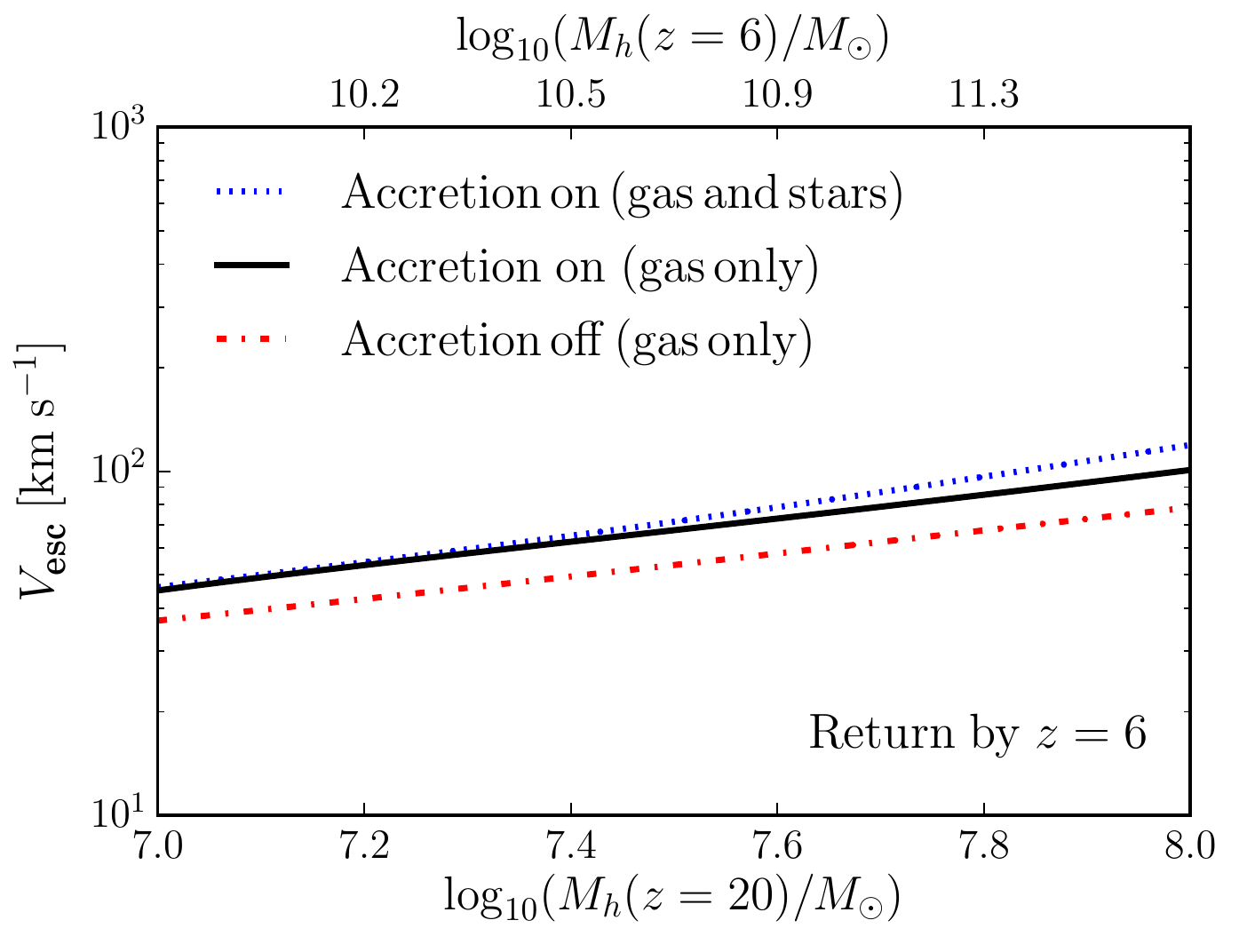}
\includegraphics[width=\columnwidth]{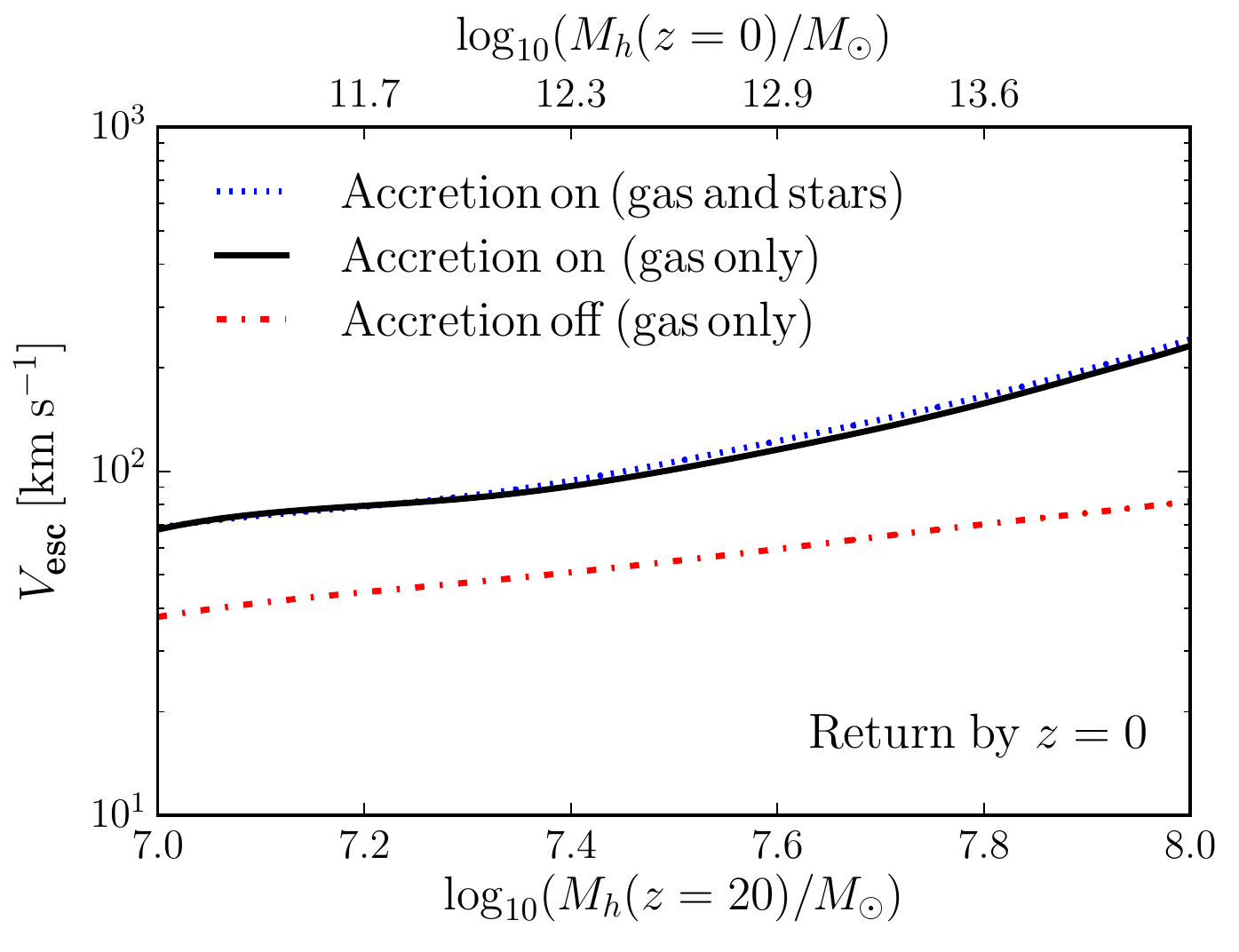}
\vspace{-5 mm}
\caption{Velocity needed to escape accreting and non-accreting halos, with return to the host required by $z=6$ (upper panel) and $z=0$ (lower panel). The kick is given at $z=20$ and stellar masses are set by the stellar mass-halo mass relation. Although highly centrally concentrated, stars do not significantly affect escape velocities because their dynamical friction effects are small at high velocities (\S \ref{subsec:results_df}).}
\label{fig:vesc}
\end{figure}

Allowing the host halo to accrete mass significantly increases SMBH escape velocities (Fig. \ref{fig:vesc}). When return is required by $z=6$, SMBH escape velocities from accreting halos are $\approx$0.1 dex higher than the case without halo accretion. When we relax this requirement such that the SMBH must return by $z=0$, the escape velocity increases by up to 0.6 dex compared to the case without halo accretion. Alternate halo accretion rates (e.g., \citealt{RP2016}) do not significantly affect these results.
\subsection{Baryon Distribution}
\label{subsec:results_baryons}
\begin{figure}
\vspace{1mm}
\includegraphics[width=\columnwidth]{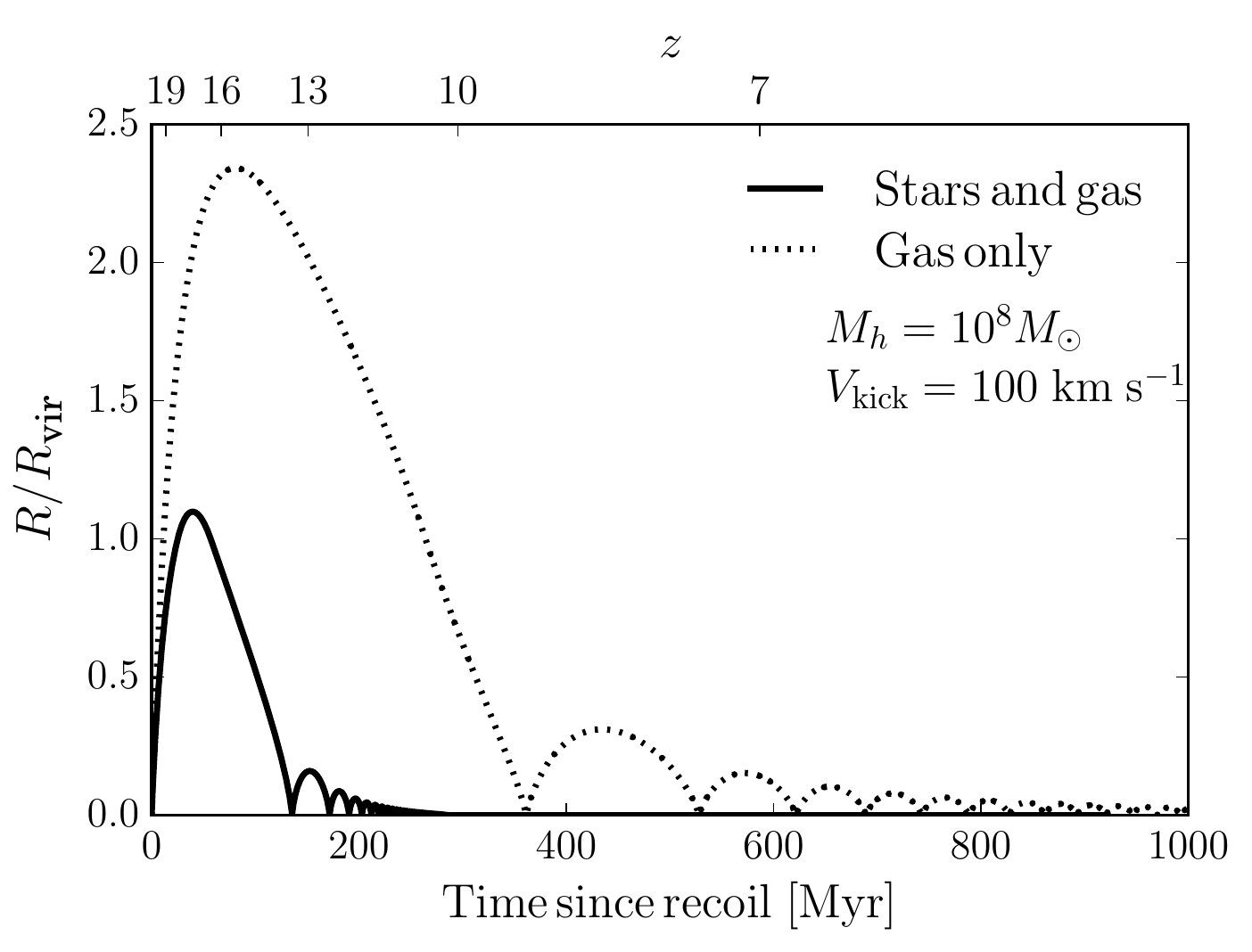}
\vspace{-6 mm}
\caption{SMBH trajectories in a purely gaseous halo and one with stars. The high stellar density at the center of the host makes escape to large radii more difficult. The kick is at $z=20$ and $\Mbh = 10^5 \Msun$.}
\label{stars_orbit}
\end{figure}
We find that the inclusion of stars in the host halo makes it harder for the SMBH to reach large radii (see Fig. \ref{stars_orbit}). This dampening results from the distribution of stars inside of the host: the stellar density is large at small radii, but drops off quickly. As a result, a significant amount of the black hole's energy can be dissipated through dynamical friction while it travels through the central regions of the halo.
The stellar profile does not, however, have a significant effect on SMBH escape velocities. As dynamical friction forces go to zero at high velocities, the central stellar densities do not significantly decrease the total energy of the black hole. 
\subsection{Seed black hole mass} 
\label{subsec:results_mbh}
\begin{figure}
\vspace{-4 mm}
\includegraphics[width=\columnwidth]{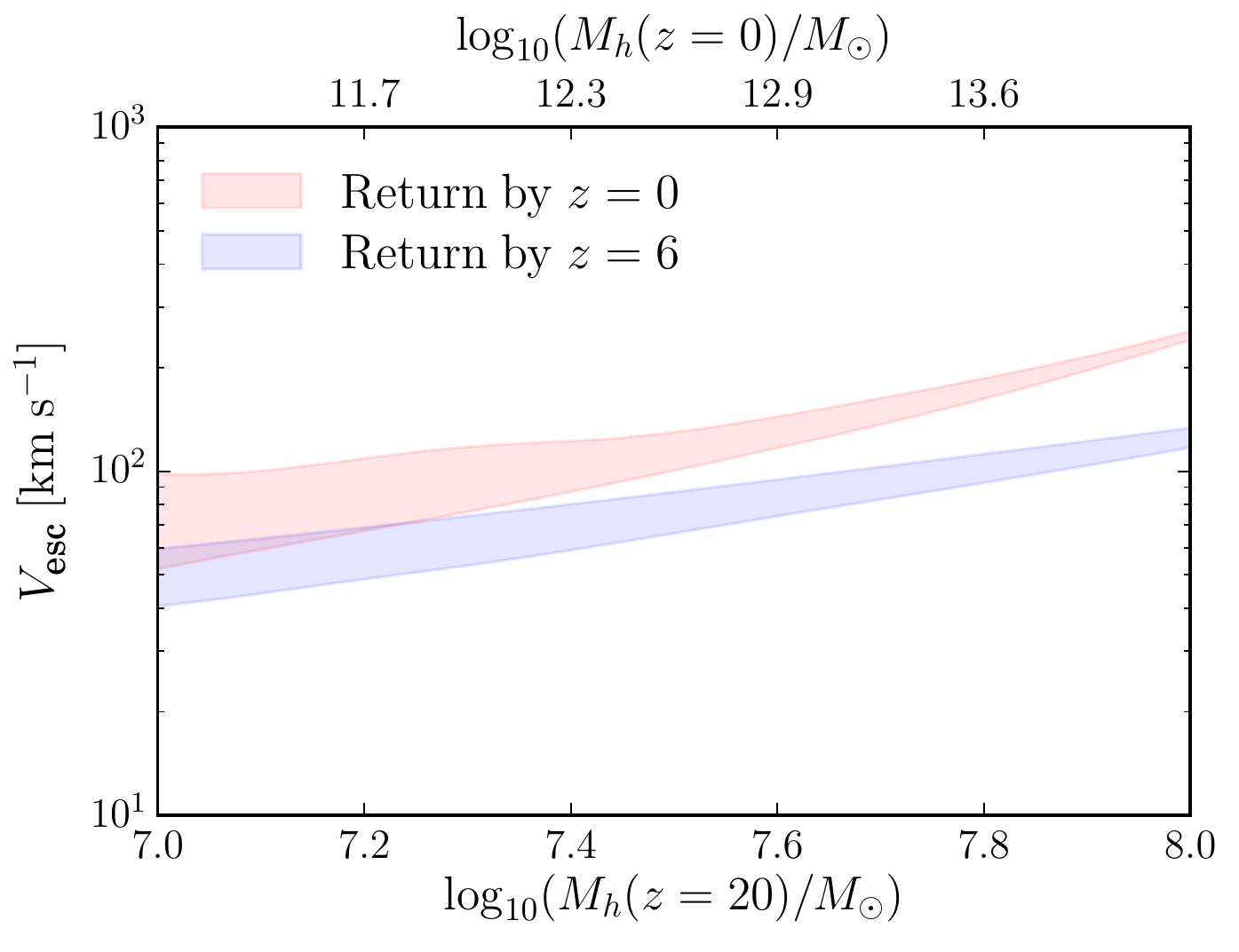}
\vspace{-4.5 mm}
\caption{The lower and upper boundaries of the shaded regions give the velocity needed to escape the halo for a $10^2 \Msun$ and $10^6 \Msun$ SMBH kicked at $z=20$ when return is required by $z=0$ (red) and $z=6$ (blue). The halo accretes mass, and stars are included. The escape velocity is not sensitive to $\Mbh$ because DF forces are small near $\Vesc$ (\S \ref{subsec:results_df}).}
\label{fig:mbh_vesc}
\end{figure}

Because the precise mechanism for the formation of seed black holes is unknown, we calculate the escape velocities at the extremes of SMBH seed masses: $10^2 \Msun$, the remnants of Population III stars, and $10^6 \Msun$, the upper limit generally placed on black holes formed through direct-collapse of pristine gas. We find the difference in escape velocities to be at most 0.27 dex for $\Mh=10^7 \Msun$. At $\Mh = 10^8 \Msun$, this difference decreases to $\lesssim 0.02$ dex (Fig. \ref{fig:mbh_vesc}). Even over 8 dex in $\Mbh$ (corresponding to $\Mbh = 10^{2} \Msun$ and $\Mbh = 10^{10} \Msun$), we find the escape velocity changes by at most 0.8 dex at $\Mh$$\sim$$10^{11} \Msun$, and decreases to a 0.15 dex change for $\Mh$$\sim$$10^{14.5} \Msun$. Regardless of the seed mass, an SMBH kicked near the escape velocity will grow negligibly. For kicks near $\Vesc$, the SMBH will spend most of its orbit at large radii, outside of the center of its host where large accretion rates are possible. As a result, $\Vesc$ decreases by only $\approx$.01 dex when BH accretion is turned off. 

The insensitivity to SMBH mass follows from the same logic that explains the insensitivity of $\Vesc$ to $\ln\Lambda$ (see Eq.\ \ref{eqn:ag_adf2}): at high velocities, dynamical friction forces are small, except for the case $\Mbh \sim \Mh$. Furthermore, when return is required by $z=0$, halo mass and accretion dominate over other effects in determining $\Vesc$.

\subsection{Host halo motions} 
\label{subsec:results_hostmove}
\begin{figure}
\vspace{-6 mm}
\includegraphics[width=\columnwidth]{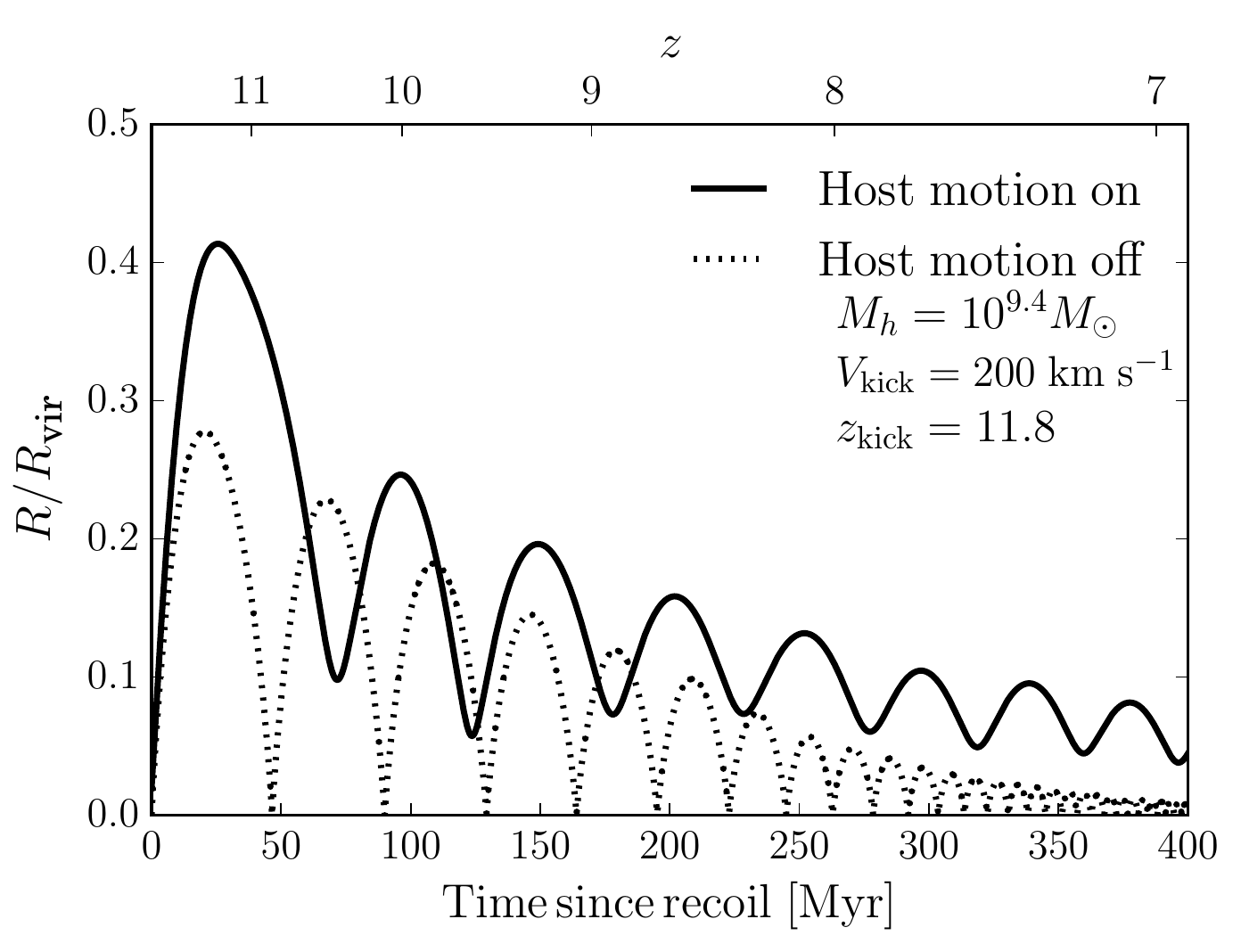}
\vspace{-6.5 mm}
\caption{Trajectories of SMBHs kicked in an example cosmologically moving and at-rest halo. Halo accretion and stars are included. The kick occurs at $z=11.8$ with $\Mbh = 10^5 \Msun$. } 
\label{fig:hm}
\end{figure}
\cite{Sijacki2011} suggested host halo motions could be a significant impediment to the return of recoiling SMBHs. We therefore allow both the host halo and the SMBH to move, following the method outlined in \textsection\ref{sec:Methodology}. Here, we follow the mass accretion histories directly from \textit{Bolshoi-Planck} rather than using the average mass accretion histories used throughout the rest of this work. Unsurprisingly, we find that a host halo that is allowed to move can significantly impact the return times of kicked SMBHs, as the SMBH must continually catch up with the host (Fig. \ref{fig:hm}). 

\begin{figure}
\vspace{2.0 mm}
\includegraphics[width=\columnwidth]{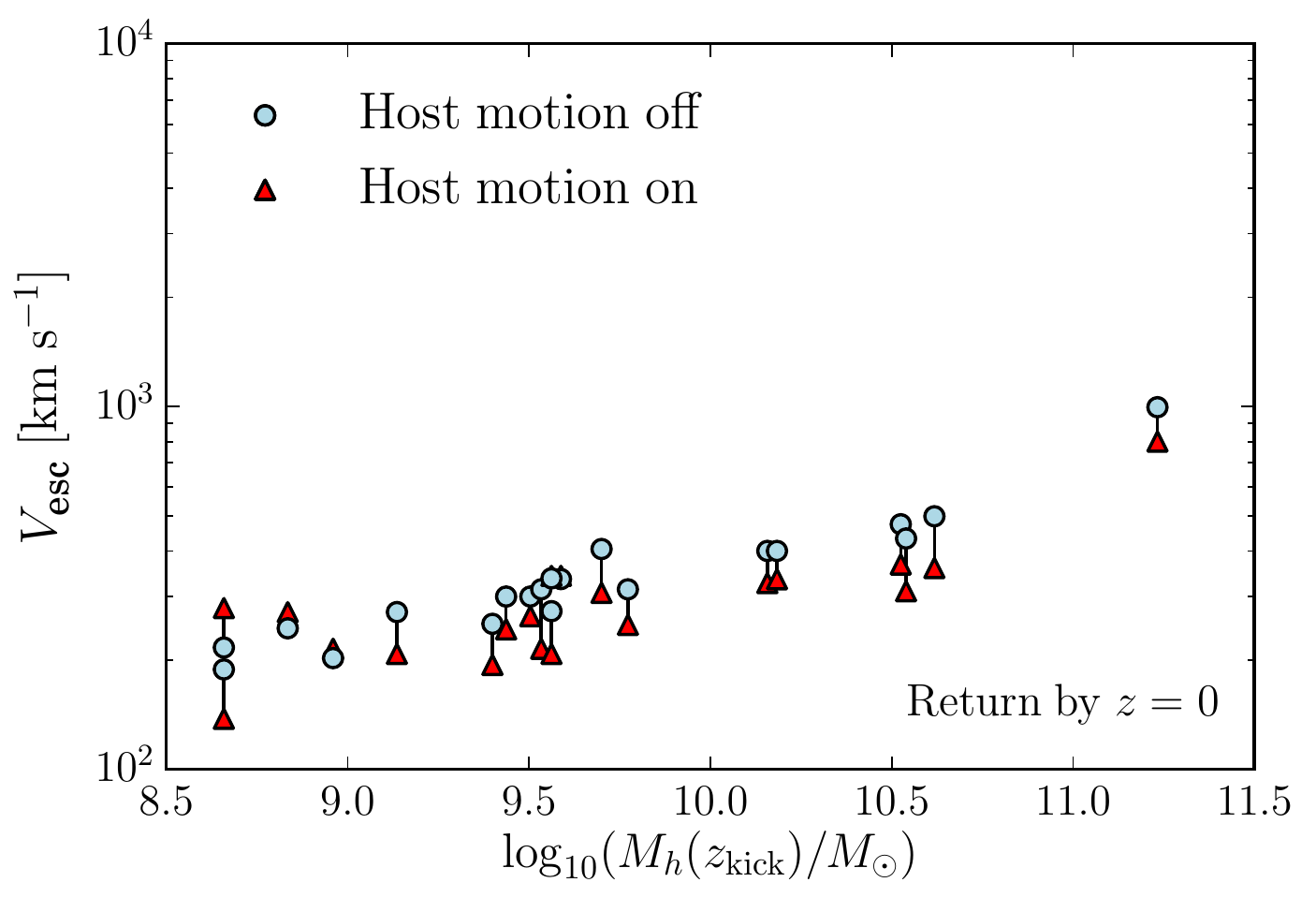}
\vspace{-3.5 mm}
\caption{Velocity needed to escape the halo, with return required by $z=0$, for several cosmologically moving and at-rest halos. The kick is given at the first appearance of the halo's main progenitor (in all cases between $z=6$ and $z=12$). The range of $z=0$ halo masses is $10^{12}\Msun < \Mh <  10^{14.5} \Msun$.} 
\label{fig:vesc_hostmotion}
\end{figure}

We also compute escape velocities for several halos  (all $>10^{12}\Msun$ at $z=0$) in \textit{Bolshoi-Planck} (Fig. \ref{fig:vesc_hostmotion}). On average, we find escape velocities only decrease by $\approx$0.05 dex. However, in some cases, if the host's motion coincides with the direction of the kick, $\Vesc$ may increase slightly. Although host motion significantly affects the shape of recoil trajectories, it has only a minor effect on escape velocities because host peculiar motions are typically much smaller than $\Vesc$ (especially at lower redshift). Instead, host motions cause the SMBH to spend significant time oscillating within $0.1\Rvir$, but at this point we consider the SMBH to have ``returned", following the criteria discussed in $\S \ref{sec:Results}$.
\subsection{Varying kick redshifts} 
\label{subsec:results_zkick}
\begin{figure}
\vspace{2 mm}
\includegraphics[width=\columnwidth]{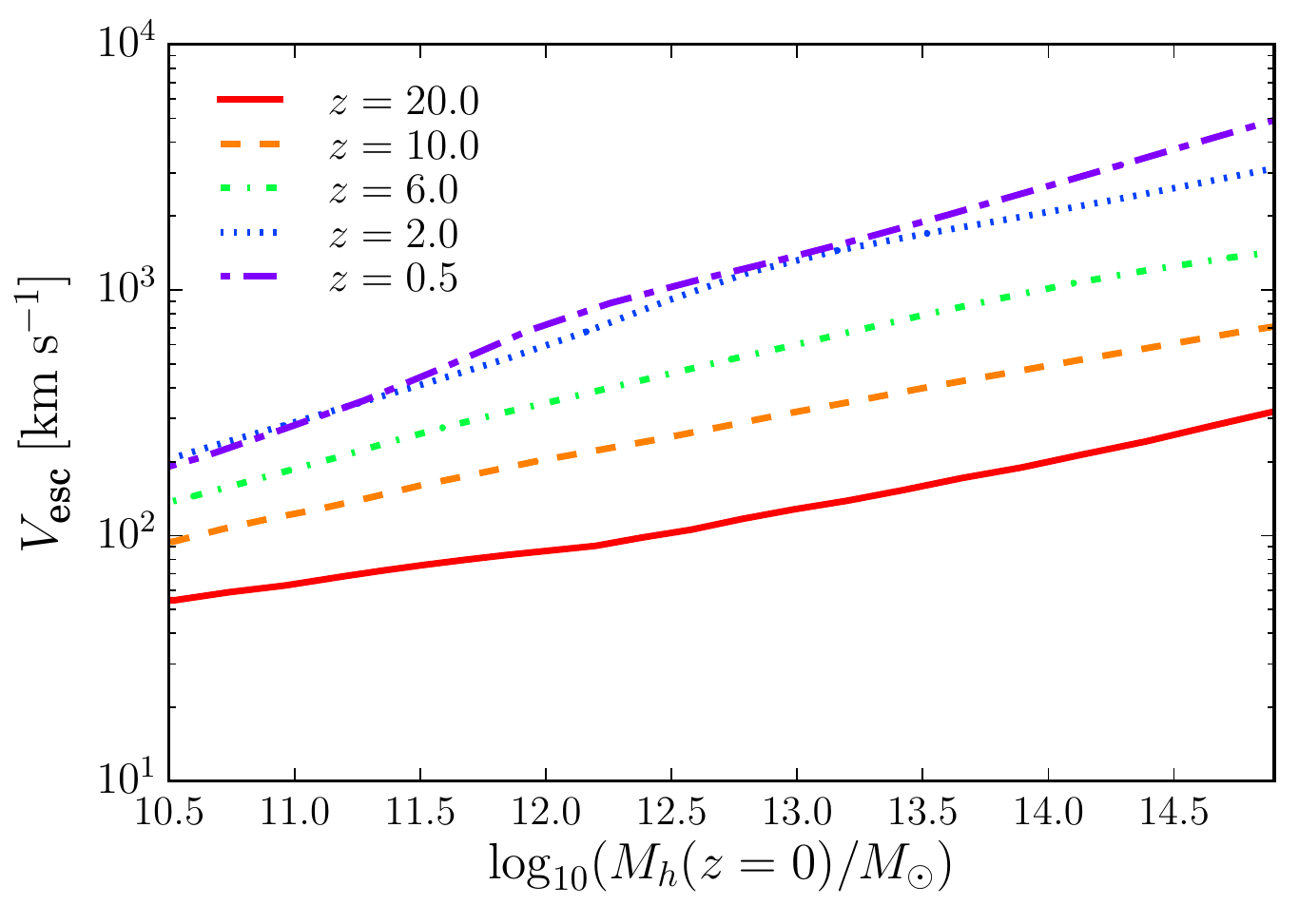}
\vspace{-6 mm}
\caption{Velocity needed to escape the halo, with return required by $z=0$, for a $10^5 \Msun$ SMBH in an accreting halo at different redshifts over a range of $z=0$ halo masses. Deviations from power-law behavior are mostly due to the shape of the stellar mass--halo mass relation. The velocity needed to escape is higher at lower redshift because the host has increased in mass.}
\label{fig:vesc_redshifts}
\end{figure}
Because recoil events can occur up to $z=0$, we examine how recoiling SMBH behavior changes with redshift. Fig. \ref{fig:vesc_redshifts} gives escape velocities over a range of $z=0$ halo masses for kicks imparted at several redshifts. It is easier for the black hole to escape at higher redshifts than at lower redshifts for the same $z=0$ halo mass because at later times the SMBH must climb out of a steeper potential well. At very low redshifts ($z \lesssim 0.5$) we observe a small downturn in the $z=0$ escape velocities. This is simply because SMBHs kicked at lower redshifts have less time to return to their host---i.e., the SMBH may be on a bound trajectory at $z=0$ but has not yet returned to within $\frac{1}{10}R_{\mathrm{vir}}$. Over all redshifts, the escape velocity follows the halo mass in a near power-law, with small deviations due to the shape of the stellar mass-halo mass relation.

We also compare escape velocities across redshifts for the same \textit{initial} host halo mass. In this case, it easiest for the SMBH to escape at lower redshifts. This is partially because of the changing definition of halo mass with redshift; a higher-redshift halo at fixed mass will have a higher circular velocity and thus a larger escape velocity.  Additionally, average halo mass accretion rates decrease monotonically with time at fixed halo mass. So, an SMBH kicked at a lower redshift moves in a potential that deepens more slowly than if the kick occurred at higher redshift.
\subsection{Formulae for escape velocities}
\label{subsec:formulae}
Here we fit escape velocities for kicked SMBHs as a function of host mass and redshift, $\Vesc(\Mh, z)$ for an accreting host halo with stellar mass set by the stellar mass-halo mass relation, the Hubble acceleration turned on, and $\ln\Lambda = 2.3$.  We set $\Mbh$ using the $M_{\mathrm{bulge}} - \Mbh$ relation given by \cite{HaringRix2004} and fit the following average $M_{\mathrm{bulge}} - M_\ast$ relation from data in \cite{Bruce2014} and \cite{Mendel2014}:
\begin{align}
\log_{10}\left(\frac{M_{\mathrm{bulge}}(M_*, a)}{\Msun}\right) = \log_{10}\left(\frac{M_*}{\Msun}\right) + \nonumber \\ 
\log_{10}\left[\frac{1 - 0.5(1-a)}{1+\exp\left(-1.13\log_{10}\left(\frac{M_*}{\Msun}\right) - 10.2\right)}\right].
\end{align}
As discussed in \S \ref{subsec:results_mbh}, changing $\Mbh$ results in extremely small corrections to escape velocities; using a different $M_{\mathrm{bulge}} - \Mbh$ relation (e.g., \citealt{mc2013} or \citealt{Kormendy2013}) or a different redshift evolution does not affect our results.

\begin{figure}
\vspace{-1 mm}
\includegraphics[width=\columnwidth]{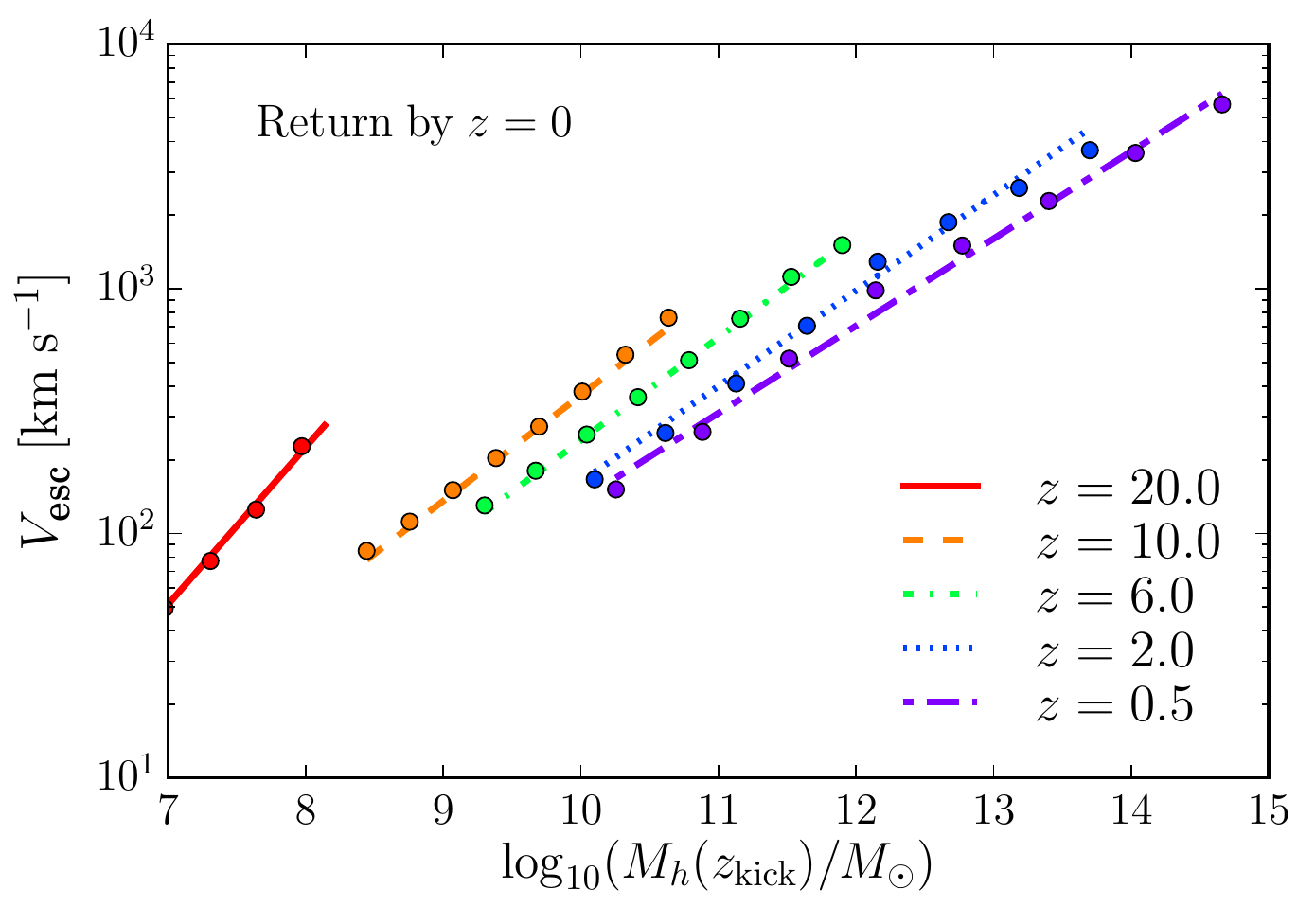}
\includegraphics[width=\columnwidth]{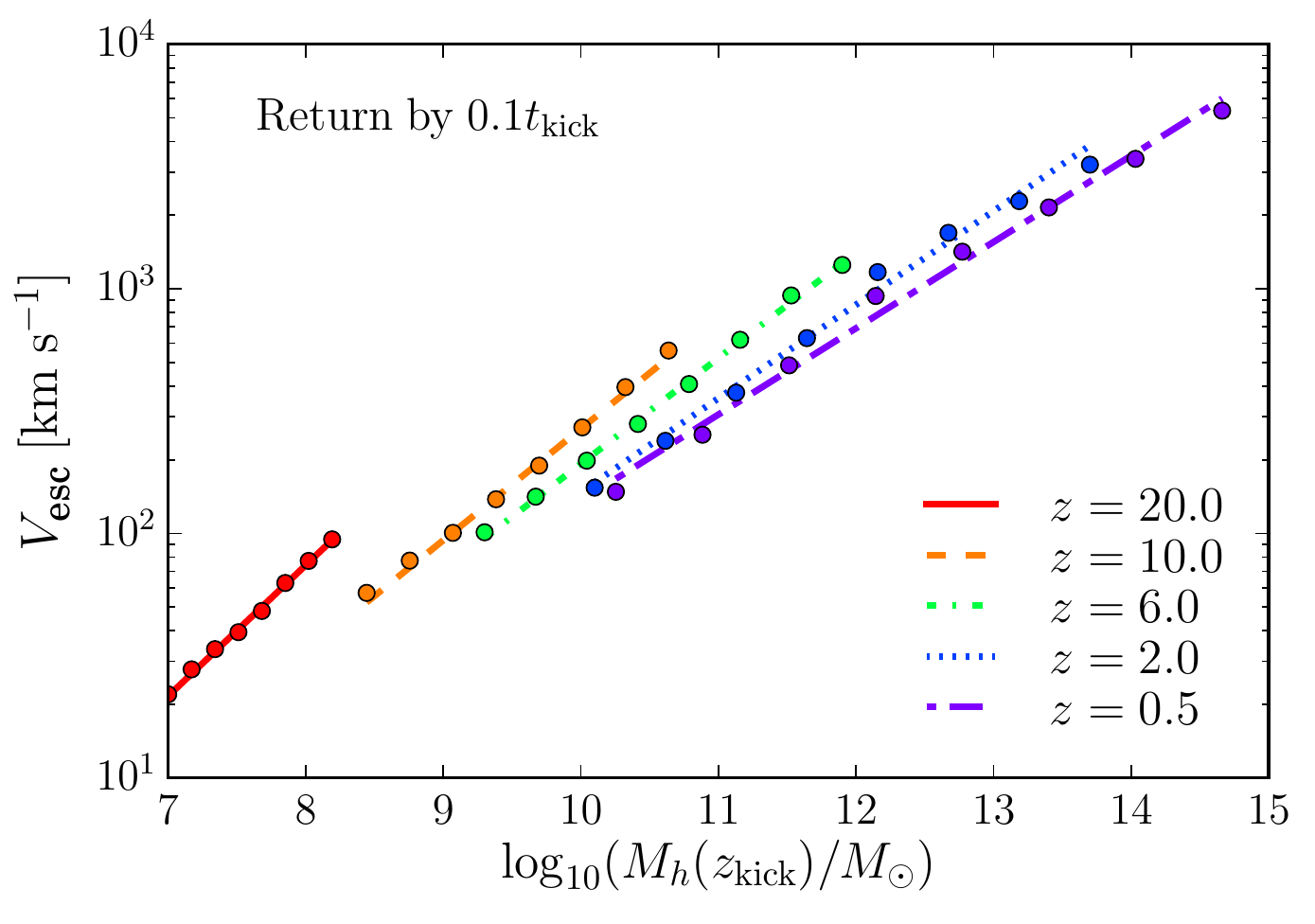}
\vspace{-4 mm}
\caption{Comparison of power-law fits (lines) and direct results from our model (scatter points) for $\Vesc(\Mh,z)$. The upper and lower panels give the velocity needed to escape the halo for return by $z=0$ (Eqs. \ref{eqn:vesc_powerlaw}-\ref{eqn:vesc_z0_fit_alpha}) and for return within 10\% of the age of the universe at the time of the kick (Eqs. \ref{eqn:vesc_powerlaw}, \ref{eqn:vesc_zdz_fit}-\ref{eqn:vesc_zdz_fit_alpha}), respectively.}
\label{fig:vesc_z0_fit}
\end{figure}

We provide fits for two definitions of $\Vesc$: return to within $0.1R_{\mathrm{vir}}$ by either $z=0$ or within $10\%$ the age of the universe at the time of the kick. We find that single power-law fits describe the host halo mass variation well:
\begin{align}
\label{eqn:vesc_powerlaw}
V_{\mathrm{esc}} (\Mh(z_{\mathrm{kick}}), z) = V_0(z)\left(\frac{\Mh(z_{\mathrm{kick}})}{10^{10} \Msun}\right)^{\alpha(z)}.
\end{align}
In both cases, $V_0(z)$ and $\alpha(z)$ are well-described by polynomials. The best fit for $z=0$ return is:
\begin{eqnarray} \label{eqn:vesc_z0_fit}
\log_{10}[V_0(z)] & = & 0.000216z^3  -0.00339z^2 \nonumber \\ && + 0.0581z + 2.10\\
\alpha(z) & = & -6.58\cdot 10^{-6}z^4 +0.000353z^3  \nonumber\\ 
&& -0.00538z^2 + 0.0342z + 0.341. \label{eqn:vesc_z0_fit_alpha}
\end{eqnarray}
For return within $\frac{1}{10}$$t_{\mathrm{kick}}$, $V_0$ and $\alpha$ evolve as:
\begin{eqnarray} \label{eqn:vesc_zdz_fit}
\log_{10}[V_0(z)] & = & 1.08\cdot 10^{-5}z^3 + 0.000710z^2 \nonumber \\ && + 0.0224z + 2.12 \\
\alpha(z) & =& 5.49\cdot 10^{-5}z^3 - 0.00183z^2 \nonumber \\ 
&& + 0.0243z + 0.341. \label{eqn:vesc_zdz_fit_alpha}
\end{eqnarray}
Comparison of these fits to our results is shown in Fig. \ref{fig:vesc_z0_fit}.
\section{Discussion \& Conclusions}
\label{sec:Conclusions}
Our main findings for supermassive black hole (SMBH) escape velocities, summarized in Table \ref{tab:conclusions}, are: 
\begin{table}
\begin{center}
\begingroup
\hspace{-10ex}\begin{tabular}{llll}
 \multicolumn{2}{|c|}{} \\
 \hline
Parameter & $\Delta V^{\mathrm{z=0}}_\mathrm{esc}$\\
 \hline
Adding halo accretion & $ \lesssim +$0.60 dex\\
Increasing $\Mbh$ from $10^{2}$ to $10^{6} \Msun$ & $\lesssim +$0.27 dex \\
Adding host halo motion & $ \lesssim$\hspace{0.1cm}-\hspace{0.05cm}0.05 
\hspace{0.02cm}dex \\
Including Hubble acceleration at high $z$ & $\lesssim +$0.01 dex \\
Increasing $\ln\Lambda$ from 2 to 4 & $\lesssim +$0.01 dex \\
Including stars & $\lesssim +$0.01 dex \\
Adding BH accretion & $\lesssim +$0.02 dex \\
\hline
\end{tabular}
\endgroup
\\
\end{center}
\caption{Effects of different parameters on escape velocities for return by $z=0$.}
\label{tab:conclusions}
\end{table}
\begin{enumerate}
\item Accretion onto the host halo significantly changes the orbits of kicked SMBHs due to the rapid increase in the mass of the host at high redshift compared to the non-accreting case. When return is required by $z=6$, the escape velocity increases by $\approx$0.1 dex. For return by $z=0$, the increase is $\approx$ 0.3 to 0.6 dex. In determining $\Vesc$, host halo accretion and mass dominate over all other factors.
\item Seed mass for SMBHs modestly affects SMBH escape velocities, with the greatest difference occurring in low mass halos. At $z=20$ and $\Mh \sim 10^{7} \Msun$ the escape velocity of $10^2$ and $10^6 \Msun$ SMBHs differs by $\approx$ 0.3 dex. 
\item SMBH trajectories are sensitive to the exact baryon distribution within the host. A host galaxy with stars damps the orbits of SMBHs due to the high central stellar densities, in agreement with \cite{Madau2004}. However, even when stars are included, SMBH escape velocities increase by $\lesssim$ 0.01 dex.
\item Cosmological motion of the host halo relative to the SMBH trajectory generally makes escape from the host easier. When the host is allowed to move it can take much longer for the SMBH to return to the host. The change in escape velocity differs between halos, but on average we find host motion leads to a decrease of $\approx$0.05 dex.
\item Including the Hubble acceleration leads to almost no changes in the orbit of the SMBH and increases escape velocities by at most $\approx$ 0.01 dex.
\item For a fixed \textit{initial} halo mass, escape from the host is easier at lower redshift because mass accretion rates decrease with cosmic time and the evolving mass definition yields a shallower potential at lower $z$ for fixed $\Mh$.  For a fixed \textit{final} halo mass, escape from the host is easier at higher redshift, because at lower redshift the host will be more massive and the SMBH must climb out of a deeper potential well.
\end{enumerate}
From these results, several observations follow:
\begin{enumerate}
\item 
The rapid mass buildup of (at least a few) SMBHs has been a challenging theoretical problem. In part, this was due to the belief that recoil events could prevent mass growth for a large fraction of the $\approx$1 Gyr between SMBH formation and $z\sim 6$ (\citealt{Haiman2004}; \citealt{Shapiro2005}; TH09). The dampening of recoil trajectories due to accretion onto the host partially mitigates this problem. SMBHs kicked in accreting halos will return to the center far quicker, enabling more continuous mass growth. 
\begin{figure}
\includegraphics[width=\columnwidth]{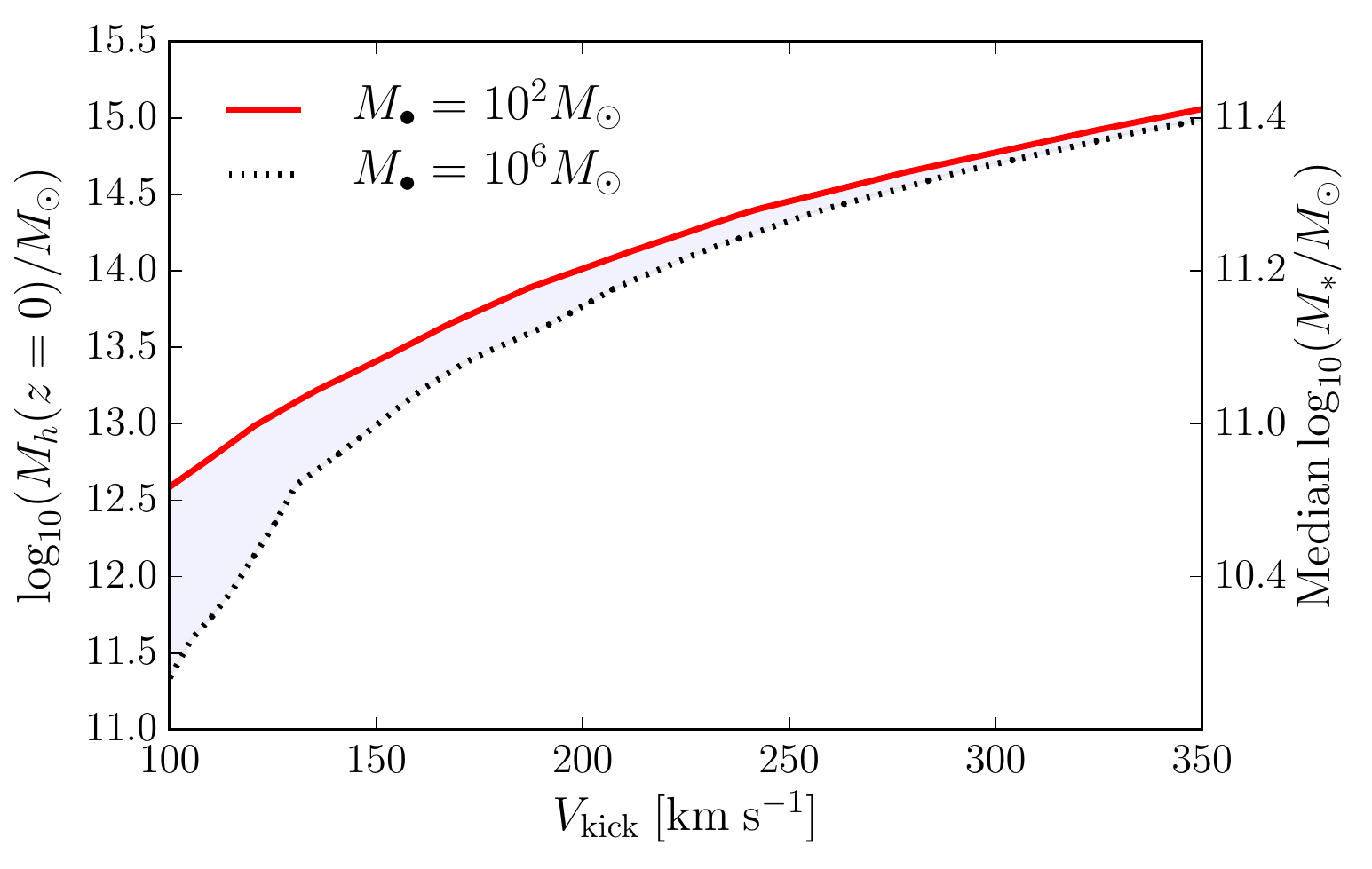}
\vspace{-4 mm}
\caption{Minimum halo mass at $z=0$ that can host an SMBH as a function of the kick velocity, assuming the central SMBH is not replenished by subsequent halo mergers, for a kick at $z=20$ to a $10^2 \Msun$ and $10^6 \Msun$ SMBH.}
\label{fig:minmass_z0}
\end{figure}
\item  
SMBH recoil velocities increase with the mass ratio between the two merging black holes, and can reach up to 3000 km s$^{-1}$ when the two black holes have randomly-oriented spins. At the very high redshifts at which seed SMBHs are hypothesized to form, such kicks are enough to permanently eject black holes from their relatively low-mass host halos (see Fig. \ref{fig:vesc_redshifts}). At $z=20$, the escape velocity from a $10^7 \Msun$ halo (which grows into a $\sim$$10^{12} \Msun$ halo by $z=0$) is $\approx$ 100 km s$^{-1}$. This problem is exacerbated if the BH seed is a $\approx 10^2 \Msun$ Population III star remnant that is constrained to sub-Eddington accretion. However, spin-aligned mergers rarely undergo kicks much larger than 300 km s$^{-1}$ \citep{Campanelli2007}. This suggests that merging black holes may have their spins aligned; if this were not the case, many mergers would lead to complete ejection from the host (even at redshifts as low as $z$$\sim$6). \cite{Bogdanovic2007} argue this scenario is possible through external torques that align the spins of the progenitors before the merger.
\item
Gravitational recoil should affect the distribution of SMBHs in galaxies \citep{RR1989}. Regardless of the exact formation time or mechanism of seed SMBHs, some small galaxies in the $z=0$ universe should lack central black holes as a result of recoils (Fig. \ref{fig:minmass_z0}). In other cases, subsequent halo mergers may replace the central SMBH. Because the halo potential will be deeper at later times, the new SMBH will be more difficult to eject (see \citealt{Volonteri2007} for SMBH occupation fraction predictions). However, small galaxies undergo fewer mergers ($\sim 0.1/\mathrm{Gyr}$ for $M_{\ast} \sim 10^8 \Msun$), and hence may retain their central SMBHs formed at early epochs \citep{casteels_etal_2014}. Observational constraints are difficult, but \cite{Miller2013} conclude that $>20\%$ of nearby early-type galaxies with $M_{\ast} < 10^{10} \Msun$ (corresponding to $\Mh \lesssim 10^{11.5} \Msun$) host a central black hole. This result is not inconsistent with our findings, but also does not verify them; further observational studies are required to probe the occupation fractions of low mass galaxies. \item 
Because smaller hosts cannot effectively keep SMBHs in their centers after a recoil, SMBHs in low mass hosts are more likely to spend significant amounts of time outside of their host's center, unable to either accrete or provide feedback to the surrounding system, thus decoupling the growth of the SMBH from that of its host. SMBHs temporarily ejected from such low mass hosts may then be captured by a nearby more massive halo \citep{Sijacki2011}. 
Unfortunately, return timescales for recoiling SMBHs are sensitive to many uncertain parameters and depend strongly on the magnitude of the recoil kick, so quantifying this effect is difficult. However, in keeping with predictions of \cite{Volonteri2007}, these results suggest increased scatter in the low mass end of the various SMBH-host galaxy scaling relations. While still controversial, \cite{mc2013} find some evidence in both the $\Mbh-\sigma$ and $\Mbh-L$ relations for this effect.
\item 
AGN luminosity functions (LFs) are available to $z\sim 6$ (\citealt{Fan2001}; \citealt{Vito2016}). If SMBH mergers are common, recoil events should cause a sharp drop in AGN luminosities at higher redshifts because recoiling SMBHs will continually be ejected from the centers of their hosts (at least until typical escape velocities are much larger than typical recoil velocities). In this case, recoiling SMBHs would be problematic for cosmic reionization via quasars, as proposed by \cite{Madau2015}. However, \cite{Volonteri2006} predict gravitational recoil to have only minor effects on the $z\sim 6$ LF because SMBH mergers in low mass hosts are rare (for further discussion see \citealt{Volonteri2007} and \citealt{Madau2004}). The Wide Field Infrared Survey Telescope (WFIRST) should be able to probe AGN luminosity functions to $z \gtrsim 6$ to test this scenario.

\item 
Off-nuclear AGN have been proposed as a possible consequence of recoiling SMBHs. Such objects are possible if the SMBH can carry its accretion disk with it as it passes through its host, and is dependent upon the amount of baryonic material available within a radius $\sim G\Mbh/V^{2}_{\mathrm{kick}}$ of the host's center. From this, \cite{Volonteri2008} predict between 1 and 30 off-nuclear AGN per deg$^2$. Using hydrodynamical simulations, \cite{Blecha2015} predict (depending on assumptions for SMBH spin alignment) between $< 1$ and $\sim 10$ per deg$^2$ offset AGN. In the past decade, there has been a growing body of evidence for the existence of off-nuclear AGN (\citealt{Komossa2008}; \citealt{Barrows2016}; \citealt{Chiaberge2016}; \citealt{Kim2016}; \citealt{Makarov2016}). 
\begin{figure}
\includegraphics[width=\columnwidth]{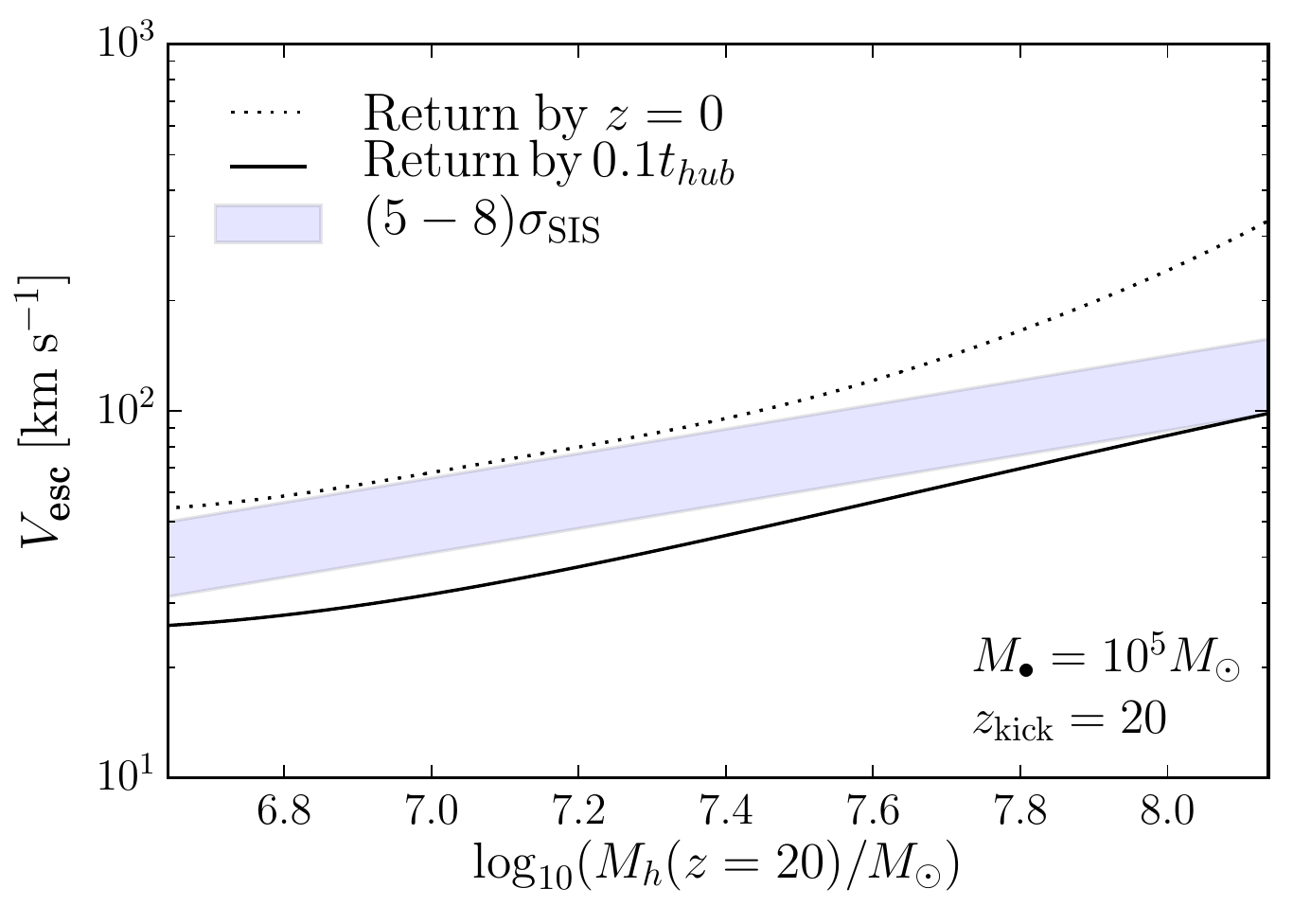}
\includegraphics[width=\columnwidth]{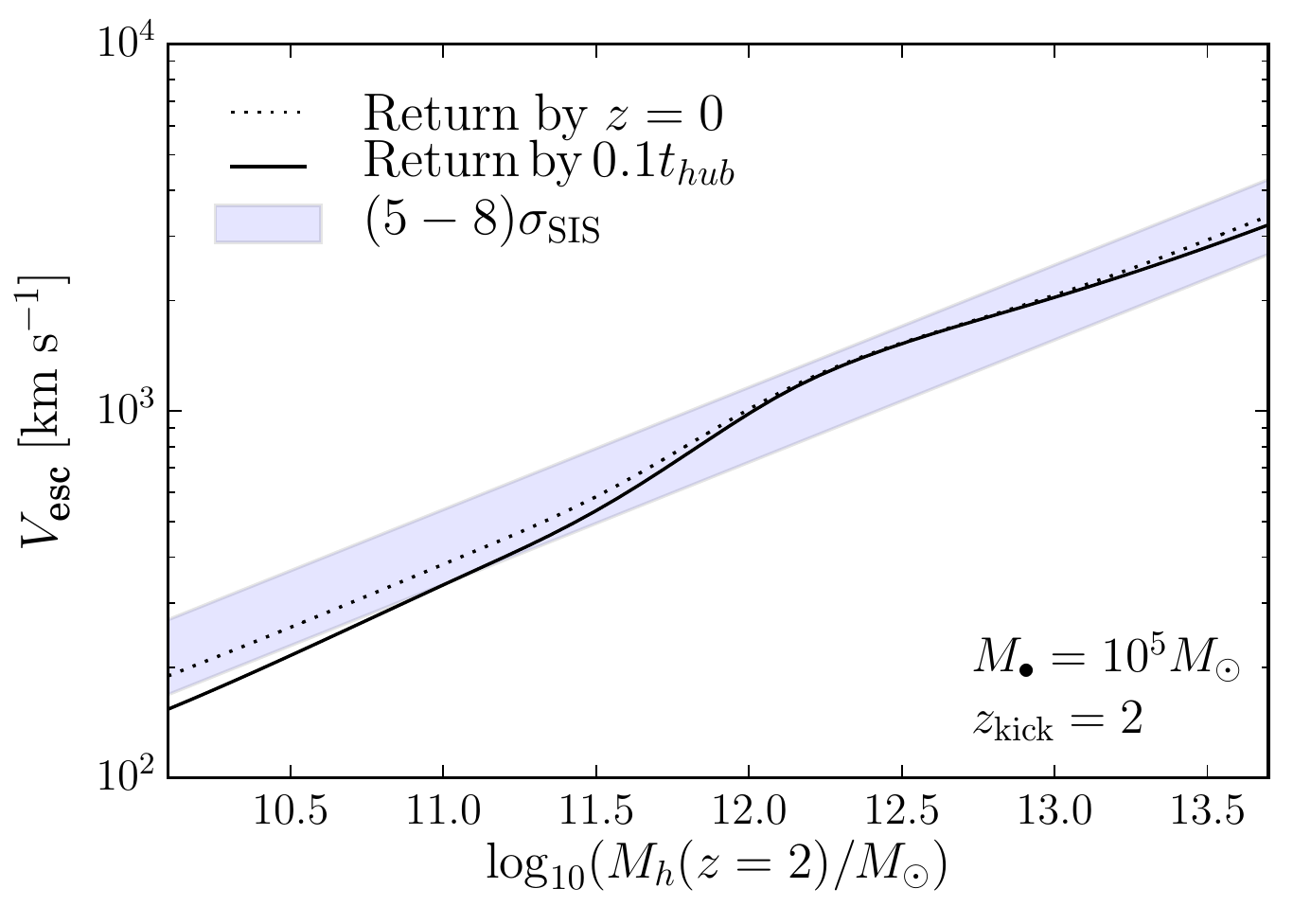}
\vspace{-6 mm}
\caption{Comparison of results from \protect\cite{Tanaka2009}, who find $\Vesc \approx (5-8) \times \sigma_{SIS}$ for return to 0.1$\Rvir$ within 10\% of the Hubble time with our calculated values. The $(5-8)\times \sigma_{SIS}$ range is an overestimate at high redshift, but agrees reasonably well with our calculations for $z \lesssim 2$.} 
\label{fig:sigma}
\end{figure}
\item
For $5 < z < 40$, TH09 find $\Vesc \approx (5-8)\times \sigma_{SIS}$, where $\sigma_{SIS} = \sqrt{G\Mh/2\Rvir}$, for return to inside $0.1\Rvir$ within 10\% of the Hubble time. For identical return criteria, we find this range overestimates $\Vesc$ by $\approx$ 0.2 dex at high redshift ($z\gtrsim$2). While TH09 do not compute $\Vesc(z < 5)$, the $(5-8)\times \sigma_{SIS}$ range is more accurate at lower redshift ($z\lesssim$2) (Fig. \ref{fig:sigma}).  As TH09 use a very similar method, we have investigated the apparent discrepancy at $z>5$. We have verified that our numerical calculations agree with the analytic solution for $\Vesc$ in the absence of dynamical friction (e.g., escape velocity of 70 km s$^{-1}$ for a non-accreting $10^8 \Msun$ halo at $z=20$); given that dynamical friction is very subdominant to gravitational forces near the escape velocity (\S \ref{ss:df}), the significant difference between the TH09 results and the analytic (no dynamical friction) solution is somewhat unexpected.
\item 
\cite{smole2015} studied recoiling SMBHs in an evolving potential using an average halo accretion rate for the specific cases of two DM-only halos with $z=0$ masses of $10^{12} \Msun$ and $2\cdot 10^{13} \Msun$ (their "Halo 1" and "Halo 2" respectively). Based on the assumption that gaseous dynamical friction forces are very strong near the center of the host halo, in their model if the SMBH passed through the center of the host it was assumed to instantaneously lose all momentum and stay there; they then define the ``critical velocity" as the kick needed such that the SMBH never returns to pass through the host's center. Using this return criterion and the same potential (i.e., DM-only, cosmologically accreting halos), we compare our computed values for the critical velocity. For Halo 1, they find $V_{\mathrm{crit}} = $300 km s$^{-1}$ and 500 km s$^{-1}$ at $z=7$ and $z=1$ respectively, in excellent agreement with our values of 305 s$^{-1}$ and 450 km s$^{-1}$. For Halo 2 they find $V_{\mathrm{crit}} = $ 725 km s$^{-1}$ and 1200 km s$^{-1}$, $\approx$ 0.1 dex larger than our values of 500  km s$^{-1}$ and 1000 km s$^{-1}$ at $z=7$ and $z=1$. The remaining differences may be due to variations in either the adopted halo concentrations or halo growth histories.
\end{enumerate}
\section*{Acknowledgements}
We thank Dan Holz and Goni Halevi for very helpful comments during the preparation of this paper. PB was supported by program number HST-HF2-51353.001-A, provided by NASA through a Hubble Fellowship grant from the Space Telescope Science Institute, which is operated by the Association of Universities for Research in Astronomy, Incorporated, under NASA contract NAS5-26555. RS acknowledges support from the 
European Research Council under the European Union's Seventh Framework Programme (FP/2007-2013) / ERC Grant Agreement n. 306476.

%%%%%%%%%%%%%%%%%%%%%%%%%%%%%%%%%%%%%%%%%%%%%%%%%%



\bibliographystyle{mnras}
\bibliography{references} 

\begin{thebibliography}{}
\makeatletter
\relax
\def\mn@urlcharsother{\let\do\@makeother \do\$\do\&\do\#\do\^\do\_\do\%\do\~}
\def\mn@doi{\begingroup\mn@urlcharsother \@ifnextchar [ {\mn@doi@}
  {\mn@doi@[]}}
\def\mn@doi@[#1]#2{\def\@tempa{#1}\ifx\@tempa\@empty \href
  {http://dx.doi.org/#2} {doi:#2}\else \href {http://dx.doi.org/#2} {#1}\fi
  \endgroup}
\def\mn@eprint#1#2{\mn@eprint@#1:#2::\@nil}
\def\mn@eprint@arXiv#1{\href {http://arxiv.org/abs/#1} {{\tt arXiv:#1}}}
\def\mn@eprint@dblp#1{\href {http://dblp.uni-trier.de/rec/bibtex/#1.xml}
  {dblp:#1}}
\def\mn@eprint@#1:#2:#3:#4\@nil{\def\@tempa {#1}\def\@tempb {#2}\def\@tempc
  {#3}\ifx \@tempc \@empty \let \@tempc \@tempb \let \@tempb \@tempa \fi \ifx
  \@tempb \@empty \def\@tempb {arXiv}\fi \@ifundefined
  {mn@eprint@\@tempb}{\@tempb:\@tempc}{\expandafter \expandafter \csname
  mn@eprint@\@tempb\endcsname \expandafter{\@tempc}}}

\bibitem[\protect\citeauthoryear{{Baker}, {Boggs}, {Centrella}, {Kelly},
  {McWilliams}, {Miller}  \& {van Meter}}{{Baker} et~al.}{2008}]{Baker2008}
{Baker} J.~G.,  {Boggs} W.~D.,  {Centrella} J.,  {Kelly} B.~J.,  {McWilliams}
  S.~T.,  {Miller} M.~C.,   {van Meter} J.~R.,  2008, \mn@doi [\apjl]
  {10.1086/590927}, \href {http://adsabs.harvard.edu/abs/2008ApJ...682L..29B}
  {682, L29}

\bibitem[\protect\citeauthoryear{{Barrows}, {Comerford}, {Greene}  \&
  {Pooley}}{{Barrows} et~al.}{2016}]{Barrows2016}
{Barrows} R.~S.,  {Comerford} J.~M.,  {Greene} J.~E.,   {Pooley} D.,  2016,
  \mn@doi [\apj] {10.3847/0004-637X/829/1/37}, \href
  {http://adsabs.harvard.edu/abs/2016ApJ...829...37B} {829, 37}

\bibitem[\protect\citeauthoryear{{Behroozi} \& {Silk}}{{Behroozi} \&
  {Silk}}{2015}]{Behroozi2015}
{Behroozi} P.~S.,  {Silk} J.,  2015, \mn@doi [\apj]
  {10.1088/0004-637X/799/1/32}, \href
  {http://adsabs.harvard.edu/abs/2015ApJ...799...32B} {799, 32}

\bibitem[\protect\citeauthoryear{{Behroozi}, {Wechsler}  \& {Wu}}{{Behroozi}
  et~al.}{2013a}]{rockstar}
{Behroozi} P.~S.,  {Wechsler} R.~H.,   {Wu} H.-Y.,  2013a, \mn@doi [\apj]
  {10.1088/0004-637X/762/2/109}, \href
  {http://adsabs.harvard.edu/abs/2013ApJ...762..109B} {762, 109}

\bibitem[\protect\citeauthoryear{{Behroozi}, {Wechsler}, {Wu}, {Busha},
  {Klypin}  \& {Primack}}{{Behroozi} et~al.}{2013b}]{ctrees}
{Behroozi} P.~S.,  {Wechsler} R.~H.,  {Wu} H.-Y.,  {Busha} M.~T.,  {Klypin}
  A.~A.,   {Primack} J.~R.,  2013b, \mn@doi [\apj]
  {10.1088/0004-637X/763/1/18}, \href
  {http://adsabs.harvard.edu/abs/2013ApJ...763...18B} {763, 18}

\bibitem[\protect\citeauthoryear{{Behroozi}, {Wechsler}  \&
  {Conroy}}{{Behroozi} et~al.}{2013c}]{Behroozi_sfh}
{Behroozi} P.~S.,  {Wechsler} R.~H.,   {Conroy} C.,  2013c, \mn@doi [\apj]
  {10.1088/0004-637X/770/1/57}, \href
  {http://adsabs.harvard.edu/abs/2013ApJ...770...57B} {770, 57}

\bibitem[\protect\citeauthoryear{{Behroozi}, {Wechsler}, {Lu}, {Hahn}, {Busha},
  {Klypin}  \& {Primack}}{{Behroozi} et~al.}{2014}]{behroozi_etal_2014}
{Behroozi} P.~S.,  {Wechsler} R.~H.,  {Lu} Y.,  {Hahn} O.,  {Busha} M.~T.,
  {Klypin} A.,   {Primack} J.~R.,  2014, \mn@doi [\apj]
  {10.1088/0004-637X/787/2/156}, \href
  {http://adsabs.harvard.edu/abs/2014ApJ...787..156B} {787, 156}

\bibitem[\protect\citeauthoryear{{Binney} \& {Tremaine}}{{Binney} \&
  {Tremaine}}{1987}]{Binney1987}
{Binney} J.,  {Tremaine} S.,  1987, {Galactic dynamics}

\bibitem[\protect\citeauthoryear{{Blecha} et~al.,}{{Blecha}
  et~al.}{2016}]{Blecha2015}
{Blecha} L.,  et~al., 2016, \mn@doi [\mnras] {10.1093/mnras/stv2646}, \href
  {http://adsabs.harvard.edu/abs/2016MNRAS.456..961B} {456, 961}

\bibitem[\protect\citeauthoryear{{Bogdanovi{\'c}}, {Reynolds}  \&
  {Miller}}{{Bogdanovi{\'c}} et~al.}{2007}]{Bogdanovic2007}
{Bogdanovi{\'c}} T.,  {Reynolds} C.~S.,   {Miller} M.~C.,  2007, \mn@doi
  [\apjl] {10.1086/518769}, \href
  {http://adsabs.harvard.edu/abs/2007ApJ...661L.147B} {661, L147}

\bibitem[\protect\citeauthoryear{{Bondi} \& {Hoyle}}{{Bondi} \&
  {Hoyle}}{1944}]{Bondi1944}
{Bondi} H.,  {Hoyle} F.,  1944, \mn@doi [\mnras] {10.1093/mnras/104.5.273},
  \href {http://adsabs.harvard.edu/abs/1944MNRAS.104..273B} {104, 273}

\bibitem[\protect\citeauthoryear{{Bromm} \& {Loeb}}{{Bromm} \&
  {Loeb}}{2003}]{Bromm2003}
{Bromm} V.,  {Loeb} A.,  2003, \mn@doi [\apj] {10.1086/377529}, \href
  {http://adsabs.harvard.edu/abs/2003ApJ...596...34B} {596, 34}

\bibitem[\protect\citeauthoryear{{Bruce} et~al.,}{{Bruce}
  et~al.}{2014}]{Bruce2014}
{Bruce} V.~A.,  et~al., 2014, \mn@doi [\mnras] {10.1093/mnras/stu1478}, \href
  {http://adsabs.harvard.edu/abs/2014MNRAS.444.1001B} {444, 1001}

\bibitem[\protect\citeauthoryear{{Bryan} \& {Norman}}{{Bryan} \&
  {Norman}}{1998}]{BN1998}
{Bryan} G.~L.,  {Norman} M.~L.,  1998, \mn@doi [\apj] {10.1086/305262}, \href
  {http://adsabs.harvard.edu/abs/1998ApJ...495...80B} {495, 80}

\bibitem[\protect\citeauthoryear{{Campanelli}, {Lousto}, {Zlochower}  \&
  {Merritt}}{{Campanelli} et~al.}{2007}]{Campanelli2007}
{Campanelli} M.,  {Lousto} C.~O.,  {Zlochower} Y.,   {Merritt} D.,  2007,
  \mn@doi [Physical Review Letters] {10.1103/PhysRevLett.98.231102}, \href
  {http://adsabs.harvard.edu/abs/2007PhRvL..98w1102C} {98, 231102}

\bibitem[\protect\citeauthoryear{{Casteels} et~al.,}{{Casteels}
  et~al.}{2014}]{casteels_etal_2014}
{Casteels} K.~R.~V.,  et~al., 2014, \mn@doi [\mnras] {10.1093/mnras/stu1799},
  \href {http://adsabs.harvard.edu/abs/2014MNRAS.445.1157C} {445, 1157}

\bibitem[\protect\citeauthoryear{{Chiaberge} et~al.,}{{Chiaberge}
  et~al.}{2016}]{Chiaberge2016}
{Chiaberge} M.,  et~al., 2016, preprint, \href
  {http://adsabs.harvard.edu/abs/2016arXiv161105501C} {} (\mn@eprint {arXiv}
  {1611.05501})

\bibitem[\protect\citeauthoryear{{De Rosa}, {Decarli}, {Walter}, {Fan},
  {Jiang}, {Kurk}, {Pasquali}  \& {Rix}}{{De Rosa} et~al.}{2011}]{DeRosa2011}
{De Rosa} G.,  {Decarli} R.,  {Walter} F.,  {Fan} X.,  {Jiang} L.,  {Kurk} J.,
  {Pasquali} A.,   {Rix} H.~W.,  2011, \mn@doi [\apj]
  {10.1088/0004-637X/739/2/56}, \href
  {http://adsabs.harvard.edu/abs/2011ApJ...739...56D} {739, 56}

\bibitem[\protect\citeauthoryear{{Diemer} \& {Kravtsov}}{{Diemer} \&
  {Kravtsov}}{2015}]{Diemer2015}
{Diemer} B.,  {Kravtsov} A.~V.,  2015, \mn@doi [\apj]
  {10.1088/0004-637X/799/1/108}, \href
  {http://adsabs.harvard.edu/abs/2015ApJ...799..108D} {799, 108}

\bibitem[\protect\citeauthoryear{{Escala}, {Larson}, {Coppi}  \&
  {Mardones}}{{Escala} et~al.}{2004}]{Escala2004}
{Escala} A.,  {Larson} R.~B.,  {Coppi} P.~S.,   {Mardones} D.,  2004, \mn@doi
  [\apj] {10.1086/386278}, \href
  {http://adsabs.harvard.edu/abs/2004ApJ...607..765E} {607, 765}

\bibitem[\protect\citeauthoryear{{Fan} et~al.,}{{Fan} et~al.}{2001}]{Fan2001}
{Fan} X.,  et~al., 2001, \mn@doi [\aj] {10.1086/324111}, \href
  {http://adsabs.harvard.edu/abs/2001AJ....122.2833F} {122, 2833}

\bibitem[\protect\citeauthoryear{{Gualandris} \& {Merritt}}{{Gualandris} \&
  {Merritt}}{2008}]{Gualandris2008}
{Gualandris} A.,  {Merritt} D.,  2008, \mn@doi [\apj] {10.1086/586877}, \href
  {http://adsabs.harvard.edu/abs/2008ApJ...678..780G} {678, 780}

\bibitem[\protect\citeauthoryear{{Haiman}}{{Haiman}}{2004}]{Haiman2004}
{Haiman} Z.,  2004, \mn@doi [\apj] {10.1086/422910}, \href
  {http://adsabs.harvard.edu/abs/2004ApJ...613...36H} {613, 36}

\bibitem[\protect\citeauthoryear{{H{\"a}ring} \& {Rix}}{{H{\"a}ring} \&
  {Rix}}{2004}]{HaringRix2004}
{H{\"a}ring} N.,  {Rix} H.-W.,  2004, \mn@doi [\apjl] {10.1086/383567}, \href
  {http://adsabs.harvard.edu/abs/2004ApJ...604L..89H} {604, L89}

\bibitem[\protect\citeauthoryear{{Heckman} \& {Kauffmann}}{{Heckman} \&
  {Kauffmann}}{2011}]{HK11}
{Heckman} T.~M.,  {Kauffmann} G.,  2011, \mn@doi [Science]
  {10.1126/science.1200504}, \href
  {http://adsabs.harvard.edu/abs/2011Sci...333..182H} {333, 182}

\bibitem[\protect\citeauthoryear{{Heger}, {Fryer}, {Woosley}, {Langer}  \&
  {Hartmann}}{{Heger} et~al.}{2003}]{Heger2003}
{Heger} A.,  {Fryer} C.~L.,  {Woosley} S.~E.,  {Langer} N.,   {Hartmann} D.~H.,
   2003, \mn@doi [\apj] {10.1086/375341}, \href
  {http://adsabs.harvard.edu/abs/2003ApJ...591..288H} {591, 288}

\bibitem[\protect\citeauthoryear{{Hernquist}}{{Hernquist}}{1990}]{hernquist}
{Hernquist} L.,  1990, \mn@doi [\apj] {10.1086/168845}, \href
  {http://adsabs.harvard.edu/abs/1990ApJ...356..359H} {356, 359}

\bibitem[\protect\citeauthoryear{{Hughes}, {Favata}  \& {Holz}}{{Hughes}
  et~al.}{2005}]{Hughes2004}
{Hughes} S.~A.,  {Favata} M.,   {Holz} D.~E.,  2005, in {Merloni} A.,
  {Nayakshin} S.,   {Sunyaev} R.~A.,  eds, Growing Black Holes: Accretion in a
  Cosmological Context. pp 333--339 (\mn@eprint {} {astro-ph/0408492}),
  \mn@doi{10.1007/11403913_64}

\bibitem[\protect\citeauthoryear{{Jiang}, {Fan}, {Vestergaard}, {Kurk},
  {Walter}, {Kelly}  \& {Strauss}}{{Jiang} et~al.}{2007}]{Jiang2007}
{Jiang} L.,  {Fan} X.,  {Vestergaard} M.,  {Kurk} J.~D.,  {Walter} F.,  {Kelly}
  B.~C.,   {Strauss} M.~A.,  2007, \mn@doi [\aj] {10.1086/520811}, \href
  {http://adsabs.harvard.edu/abs/2007AJ....134.1150J} {134, 1150}

\bibitem[\protect\citeauthoryear{{Johnson} \& {Haardt}}{{Johnson} \&
  {Haardt}}{2016}]{Johnson2016}
{Johnson} J.~L.,  {Haardt} F.,  2016, \mn@doi [\pasa] {10.1017/pasa.2016.4},
  \href {http://adsabs.harvard.edu/abs/2016PASA...33....7J} {33, e007}

\bibitem[\protect\citeauthoryear{{Kim}, {Evans}, {Stierwalt}  \&
  {Privon}}{{Kim} et~al.}{2016}]{Kim2016}
{Kim} D.-C.,  {Evans} A.~S.,  {Stierwalt} S.,   {Privon} G.~C.,  2016, \mn@doi
  [\apj] {10.3847/0004-637X/824/2/122}, \href
  {http://adsabs.harvard.edu/abs/2016ApJ...824..122K} {824, 122}

\bibitem[\protect\citeauthoryear{{Klypin}, {Trujillo-Gomez}  \&
  {Primack}}{{Klypin} et~al.}{2011}]{Klypin2010}
{Klypin} A.~A.,  {Trujillo-Gomez} S.,   {Primack} J.,  2011, \mn@doi [\apj]
  {10.1088/0004-637X/740/2/102}, \href
  {http://adsabs.harvard.edu/abs/2011ApJ...740..102K} {740, 102}

\bibitem[\protect\citeauthoryear{{Klypin}, {Yepes}, {Gottl{\"o}ber}, {Prada}
  \& {He{\ss}}}{{Klypin} et~al.}{2016}]{Klypin2016}
{Klypin} A.,  {Yepes} G.,  {Gottl{\"o}ber} S.,  {Prada} F.,   {He{\ss}} S.,
  2016, \mn@doi [\mnras] {10.1093/mnras/stw248}, \href
  {http://adsabs.harvard.edu/abs/2016MNRAS.457.4340K} {457, 4340}

\bibitem[\protect\citeauthoryear{{Komossa}, {Zhou}  \& {Lu}}{{Komossa}
  et~al.}{2008}]{Komossa2008}
{Komossa} S.,  {Zhou} H.,   {Lu} H.,  2008, \mn@doi [\apjl] {10.1086/588656},
  \href {http://adsabs.harvard.edu/abs/2008ApJ...678L..81K} {678, L81}

\bibitem[\protect\citeauthoryear{{Kormendy} \& {Ho}}{{Kormendy} \&
  {Ho}}{2013}]{Kormendy2013}
{Kormendy} J.,  {Ho} L.~C.,  2013, \mn@doi [\araa]
  {10.1146/annurev-astro-082708-101811}, \href
  {http://adsabs.harvard.edu/abs/2013ARA%26A..51..511K} {51, 511}

\bibitem[\protect\citeauthoryear{{Latif} \& {Ferrara}}{{Latif} \&
  {Ferrara}}{2016}]{Latif2016}
{Latif} M.~A.,  {Ferrara} A.,  2016, \mn@doi [\pasa] {10.1017/pasa.2016.41},
  \href {http://adsabs.harvard.edu/abs/2016PASA...33...51L} {33, e051}

\bibitem[\protect\citeauthoryear{{Machacek}, {Bryan}  \& {Abel}}{{Machacek}
  et~al.}{2001}]{Machacek2001}
{Machacek} M.~E.,  {Bryan} G.~L.,   {Abel} T.,  2001, \mn@doi [\apj]
  {10.1086/319014}, \href {http://adsabs.harvard.edu/abs/2001ApJ...548..509M}
  {548, 509}

\bibitem[\protect\citeauthoryear{{Madau} \& {Dickinson}}{{Madau} \&
  {Dickinson}}{2014}]{MD2014}
{Madau} P.,  {Dickinson} M.,  2014, \mn@doi [\araa]
  {10.1146/annurev-astro-081811-125615}, \href
  {http://adsabs.harvard.edu/abs/2014ARA%26A..52..415M} {52, 415}

\bibitem[\protect\citeauthoryear{{Madau} \& {Haardt}}{{Madau} \&
  {Haardt}}{2015}]{Madau2015}
{Madau} P.,  {Haardt} F.,  2015, \mn@doi [\apjl] {10.1088/2041-8205/813/1/L8},
  \href {http://adsabs.harvard.edu/abs/2015ApJ...813L...8M} {813, L8}

\bibitem[\protect\citeauthoryear{{Madau} \& {Quataert}}{{Madau} \&
  {Quataert}}{2004}]{Madau2004}
{Madau} P.,  {Quataert} E.,  2004, \mn@doi [\apjl] {10.1086/421017}, \href
  {http://adsabs.harvard.edu/abs/2004ApJ...606L..17M} {606, L17}

\bibitem[\protect\citeauthoryear{{Madau}, {Haardt}  \& {Dotti}}{{Madau}
  et~al.}{2014}]{Madau2014}
{Madau} P.,  {Haardt} F.,   {Dotti} M.,  2014, \mn@doi [\apjl]
  {10.1088/2041-8205/784/2/L38}, \href
  {http://adsabs.harvard.edu/abs/2014ApJ...784L..38M} {784, L38}

\bibitem[\protect\citeauthoryear{{Makarov}, {Frouard}, {Berghea}, {Rest},
  {Chambers}, {Kaiser}, {Kudritzki}  \& {Magnier}}{{Makarov}
  et~al.}{2017}]{Makarov2016}
{Makarov} V.~V.,  {Frouard} J.,  {Berghea} C.~T.,  {Rest} A.,  {Chambers}
  K.~C.,  {Kaiser} N.,  {Kudritzki} R.-P.,   {Magnier} E.~A.,  2017, \mn@doi
  [\apjl] {10.3847/2041-8213/835/2/L30}, \href
  {http://adsabs.harvard.edu/abs/2017ApJ...835L..30M} {835, L30}

\bibitem[\protect\citeauthoryear{{McConnell} \& {Ma}}{{McConnell} \&
  {Ma}}{2013}]{mc2013}
{McConnell} N.~J.,  {Ma} C.-P.,  2013, \mn@doi [\apj]
  {10.1088/0004-637X/764/2/184}, \href
  {http://adsabs.harvard.edu/abs/2013ApJ...764..184M} {764, 184}

\bibitem[\protect\citeauthoryear{{Mendel}, {Simard}, {Palmer}, {Ellison}  \&
  {Patton}}{{Mendel} et~al.}{2014}]{Mendel2014}
{Mendel} J.~T.,  {Simard} L.,  {Palmer} M.,  {Ellison} S.~L.,   {Patton} D.~R.,
   2014, \mn@doi [\apjs] {10.1088/0067-0049/210/1/3}, \href
  {http://adsabs.harvard.edu/abs/2014ApJS..210....3M} {210, 3}

\bibitem[\protect\citeauthoryear{{Miller}, {Gallo}, {Greene}, {Kelly}, {Treu},
  {Woo}  \& {Baldassare}}{{Miller} et~al.}{2015}]{Miller2013}
{Miller} B.~P.,  {Gallo} E.,  {Greene} J.~E.,  {Kelly} B.~C.,  {Treu} T.,
  {Woo} J.-H.,   {Baldassare} V.,  2015, \mn@doi [\apj]
  {10.1088/0004-637X/799/1/98}, \href
  {http://adsabs.harvard.edu/abs/2015ApJ...799...98M} {799, 98}

\bibitem[\protect\citeauthoryear{{Milosavljevi{\'c}} \&
  {Merritt}}{{Milosavljevi{\'c}} \& {Merritt}}{2003}]{Milosavljevic2003}
{Milosavljevi{\'c}} M.,  {Merritt} D.,  2003, in {Centrella} J.~M.,  ed.,
  American Institute of Physics Conference Series Vol. 686, The Astrophysics of
  Gravitational Wave Sources. pp 201--210 (\mn@eprint {} {astro-ph/0212270}),
  \mn@doi{10.1063/1.1629432}

\bibitem[\protect\citeauthoryear{{Mortlock} et~al.,}{{Mortlock}
  et~al.}{2011}]{Mortlock2011}
{Mortlock} D.~J.,  et~al., 2011, \mn@doi [\nat] {10.1038/nature10159}, \href
  {http://adsabs.harvard.edu/abs/2011Natur.474..616M} {474, 616}

\bibitem[\protect\citeauthoryear{{Moster}, {Macci{\`o}}, {Somerville}, {Naab}
  \& {Cox}}{{Moster} et~al.}{2012}]{moster_etal_2012}
{Moster} B.~P.,  {Macci{\`o}} A.~V.,  {Somerville} R.~S.,  {Naab} T.,   {Cox}
  T.~J.,  2012, \mn@doi [\mnras] {10.1111/j.1365-2966.2012.20915.x}, \href
  {http://adsabs.harvard.edu/abs/2012MNRAS.423.2045M} {423, 2045}

\bibitem[\protect\citeauthoryear{{Nandra}, {Lasenby}  \& {Hobson}}{{Nandra}
  et~al.}{2012}]{Nandra2012}
{Nandra} R.,  {Lasenby} A.~N.,   {Hobson} M.~P.,  2012, \mn@doi [\mnras]
  {10.1111/j.1365-2966.2012.20618.x}, \href
  {http://adsabs.harvard.edu/abs/2012MNRAS.422.2931N} {422, 2931}

\bibitem[\protect\citeauthoryear{{Navarro}, {Frenk}  \& {White}}{{Navarro}
  et~al.}{1997}]{Navarro1997}
{Navarro} J.~F.,  {Frenk} C.~S.,   {White} S.~D.~M.,  1997, \apj, \href
  {http://adsabs.harvard.edu/abs/1997ApJ...490..493N} {490, 493}

\bibitem[\protect\citeauthoryear{{Ohsuga} \& {Mineshige}}{{Ohsuga} \&
  {Mineshige}}{2007}]{Ohsuga2007}
{Ohsuga} K.,  {Mineshige} S.,  2007, \mn@doi [\apj] {10.1086/522324}, \href
  {http://adsabs.harvard.edu/abs/2007ApJ...670.1283O} {670, 1283}

\bibitem[\protect\citeauthoryear{{Omukai}, {Schneider}  \& {Haiman}}{{Omukai}
  et~al.}{2008}]{Omukai2007}
{Omukai} K.,  {Schneider} R.,   {Haiman} Z.,  2008, \mn@doi [\apj]
  {10.1086/591636}, \href {http://adsabs.harvard.edu/abs/2008ApJ...686..801O}
  {686, 801}

\bibitem[\protect\citeauthoryear{{Ostriker}}{{Ostriker}}{1999}]{Ostriker1999}
{Ostriker} E.~C.,  1999, \mn@doi [\apj] {10.1086/306858}, \href
  {http://adsabs.harvard.edu/abs/1999ApJ...513..252O} {513, 252}

\bibitem[\protect\citeauthoryear{{Pezzulli}, {Valiante}  \&
  {Schneider}}{{Pezzulli} et~al.}{2016}]{Pezzulli2016}
{Pezzulli} E.,  {Valiante} R.,   {Schneider} R.,  2016, \mn@doi [\mnras]
  {10.1093/mnras/stw505}, \href
  {http://adsabs.harvard.edu/abs/2016MNRAS.458.3047P} {458, 3047}

\bibitem[\protect\citeauthoryear{{Planck Collaboration} et~al.,}{{Planck
  Collaboration} et~al.}{2015}]{Planck2015}
{Planck Collaboration} et~al., 2015, \mn@doi [\aap]
  {10.1051/0004-6361/201525830}, \href
  {http://adsabs.harvard.edu/abs/2016A%26A...594A..13P} {594, A13}

\bibitem[\protect\citeauthoryear{{Redmount} \& {Rees}}{{Redmount} \&
  {Rees}}{1989}]{RR1989}
{Redmount} I.~H.,  {Rees} M.~J.,  1989, Comments on Astrophysics, \href
  {http://adsabs.harvard.edu/abs/1989ComAp..14..165R} {14, 165}

\bibitem[\protect\citeauthoryear{{Rodr{\'{\i}}guez-Puebla}, {Behroozi},
  {Primack}, {Klypin}, {Lee}  \& {Hellinger}}{{Rodr{\'{\i}}guez-Puebla}
  et~al.}{2016a}]{Bolshoi}
{Rodr{\'{\i}}guez-Puebla} A.,  {Behroozi} P.,  {Primack} J.,  {Klypin} A.,
  {Lee} C.,   {Hellinger} D.,  2016a, \mn@doi [\mnras] {10.1093/mnras/stw1705},
  \href {http://adsabs.harvard.edu/abs/2016MNRAS.462..893R} {462, 893}

\bibitem[\protect\citeauthoryear{{Rodr{\'{\i}}guez-Puebla}, {Behroozi},
  {Primack}, {Klypin}, {Lee}  \& {Hellinger}}{{Rodr{\'{\i}}guez-Puebla}
  et~al.}{2016b}]{RP2016}
{Rodr{\'{\i}}guez-Puebla} A.,  {Behroozi} P.,  {Primack} J.,  {Klypin} A.,
  {Lee} C.,   {Hellinger} D.,  2016b, \mn@doi [\mnras] {10.1093/mnras/stw1705},
  \href {http://adsabs.harvard.edu/abs/2016MNRAS.462..893R} {462, 893}

\bibitem[\protect\citeauthoryear{{Schindler}, {Fan}  \& {Duschl}}{{Schindler}
  et~al.}{2016}]{Schindler2016}
{Schindler} J.-T.,  {Fan} X.,   {Duschl} W.~J.,  2016, \mn@doi [\apj]
  {10.3847/0004-637X/826/1/67}, \href
  {http://adsabs.harvard.edu/abs/2016ApJ...826...67S} {826, 67}

\bibitem[\protect\citeauthoryear{{Shang}, {Bryan}  \& {Haiman}}{{Shang}
  et~al.}{2010}]{Shang2009}
{Shang} C.,  {Bryan} G.~L.,   {Haiman} Z.,  2010, \mn@doi [\mnras]
  {10.1111/j.1365-2966.2009.15960.x}, \href
  {http://adsabs.harvard.edu/abs/2010MNRAS.402.1249S} {402, 1249}

\bibitem[\protect\citeauthoryear{{Shapiro}}{{Shapiro}}{2005}]{Shapiro2005}
{Shapiro} S.~L.,  2005, \mn@doi [\apj] {10.1086/427065}, \href
  {http://adsabs.harvard.edu/abs/2005ApJ...620...59S} {620, 59}

\bibitem[\protect\citeauthoryear{{Sijacki}, {Springel}  \&
  {Haehnelt}}{{Sijacki} et~al.}{2011}]{Sijacki2011}
{Sijacki} D.,  {Springel} V.,   {Haehnelt} M.~G.,  2011, \mn@doi [\mnras]
  {10.1111/j.1365-2966.2011.18666.x}, \href
  {http://adsabs.harvard.edu/abs/2011MNRAS.414.3656S} {414, 3656}

\bibitem[\protect\citeauthoryear{{Smole}}{{Smole}}{2015}]{smole2015}
{Smole} M.,  2015, \mn@doi [Serbian Astronomical Journal]
  {10.2298/SAJ150706001S}, \href
  {http://adsabs.harvard.edu/abs/2015SerAJ.191...17S} {191, 17}

\bibitem[\protect\citeauthoryear{{Somerville} \& {Dav{\'e}}}{{Somerville} \&
  {Dav{\'e}}}{2015}]{SomDave2015}
{Somerville} R.~S.,  {Dav{\'e}} R.,  2015, \mn@doi [\araa]
  {10.1146/annurev-astro-082812-140951}, \href
  {http://adsabs.harvard.edu/abs/2015ARA%26A..53...51S} {53, 51}

\bibitem[\protect\citeauthoryear{{Somerville} et~al.,}{{Somerville}
  et~al.}{2017}]{somerville_etal_2014}
{Somerville} R.~S.,  et~al., 2017, preprint, \href
  {http://adsabs.harvard.edu/abs/2017arXiv170103526S} {} (\mn@eprint {arXiv}
  {1701.03526})

\bibitem[\protect\citeauthoryear{{Tanaka} \& {Haiman}}{{Tanaka} \&
  {Haiman}}{2009}]{Tanaka2009}
{Tanaka} T.,  {Haiman} Z.,  2009, \mn@doi [\apj]
  {10.1088/0004-637X/696/2/1798}, \href
  {http://adsabs.harvard.edu/abs/2009ApJ...696.1798T} {696, 1798}

\bibitem[\protect\citeauthoryear{{Tremmel}, {Governato}, {Volonteri}  \&
  {Quinn}}{{Tremmel} et~al.}{2015}]{Tremmel2015}
{Tremmel} M.,  {Governato} F.,  {Volonteri} M.,   {Quinn} T.~R.,  2015, \mn@doi
  [\mnras] {10.1093/mnras/stv1060}, \href
  {http://adsabs.harvard.edu/abs/2015MNRAS.451.1868T} {451, 1868}

\bibitem[\protect\citeauthoryear{{Valiante}, {Schneider}, {Volonteri}  \&
  {Omukai}}{{Valiante} et~al.}{2016}]{Valiante2016}
{Valiante} R.,  {Schneider} R.,  {Volonteri} M.,   {Omukai} K.,  2016, \mn@doi
  [\mnras] {10.1093/mnras/stw225}, \href
  {http://adsabs.harvard.edu/abs/2016MNRAS.457.3356V} {457, 3356}

\bibitem[\protect\citeauthoryear{{Valiante}, {Agarwal}, {Habouzit}  \&
  {Pezzulli}}{{Valiante} et~al.}{2017}]{Valiante2017}
{Valiante} R.,  {Agarwal} B.,  {Habouzit} M.,   {Pezzulli} E.,  2017, preprint,
  \href {http://adsabs.harvard.edu/abs/2017arXiv170303808V} {} (\mn@eprint
  {arXiv} {1703.03808})

\bibitem[\protect\citeauthoryear{{Vito} et~al.,}{{Vito}
  et~al.}{2016}]{Vito2016}
{Vito} F.,  et~al., 2016, \mn@doi [\mnras] {10.1093/mnras/stw1998}, \href
  {http://adsabs.harvard.edu/abs/2016MNRAS.463..348V} {463, 348}

\bibitem[\protect\citeauthoryear{{Volonteri}}{{Volonteri}}{2007}]{Volonteri2007}
{Volonteri} M.,  2007, \mn@doi [\apjl] {10.1086/519525}, \href
  {http://adsabs.harvard.edu/abs/2007ApJ...663L...5V} {663, L5}

\bibitem[\protect\citeauthoryear{{Volonteri}}{{Volonteri}}{2012}]{Volonteri_mbh_review}
{Volonteri} M.,  2012, \mn@doi [Science] {10.1126/science.1220843}, \href
  {http://adsabs.harvard.edu/abs/2012Sci...337..544V} {337, 544}

\bibitem[\protect\citeauthoryear{{Volonteri} \& {Madau}}{{Volonteri} \&
  {Madau}}{2008}]{Volonteri2008}
{Volonteri} M.,  {Madau} P.,  2008, \mn@doi [\apjl] {10.1086/593353}, \href
  {http://adsabs.harvard.edu/abs/2008ApJ...687L..57V} {687, L57}

\bibitem[\protect\citeauthoryear{{Volonteri} \& {Rees}}{{Volonteri} \&
  {Rees}}{2005}]{VR2005}
{Volonteri} M.,  {Rees} M.~J.,  2005, \mn@doi [\apj] {10.1086/466521}, \href
  {http://adsabs.harvard.edu/abs/2005ApJ...633..624V} {633, 624}

\bibitem[\protect\citeauthoryear{{Volonteri} \& {Rees}}{{Volonteri} \&
  {Rees}}{2006}]{Volonteri2006}
{Volonteri} M.,  {Rees} M.~J.,  2006, \mn@doi [\apj] {10.1086/507444}, \href
  {http://adsabs.harvard.edu/abs/2006ApJ...650..669V} {650, 669}

\bibitem[\protect\citeauthoryear{{Volonteri}, {G{\"u}ltekin}  \&
  {Dotti}}{{Volonteri} et~al.}{2010}]{Volonteri2010}
{Volonteri} M.,  {G{\"u}ltekin} K.,   {Dotti} M.,  2010, \mn@doi [\mnras]
  {10.1111/j.1365-2966.2010.16431.x}, \href
  {http://adsabs.harvard.edu/abs/2010MNRAS.404.2143V} {404, 2143}

\bibitem[\protect\citeauthoryear{{Volonteri}, {Silk}  \& {Dubus}}{{Volonteri}
  et~al.}{2015}]{volonteri_superedd}
{Volonteri} M.,  {Silk} J.,   {Dubus} G.,  2015, \mn@doi [\apj]
  {10.1088/0004-637X/804/2/148}, \href
  {http://adsabs.harvard.edu/abs/2015ApJ...804..148V} {804, 148}

\bibitem[\protect\citeauthoryear{{Wang} et~al.,}{{Wang}
  et~al.}{2015}]{Wang2015}
{Wang} F.,  et~al., 2015, \mn@doi [\apjl] {10.1088/2041-8205/807/1/L9}, \href
  {http://adsabs.harvard.edu/abs/2015ApJ...807L...9W} {807, L9}

\bibitem[\protect\citeauthoryear{{Wu} et~al.,}{{Wu} et~al.}{2015}]{Wu2015}
{Wu} X.-B.,  et~al., 2015, \mn@doi [\nat] {10.1038/nature14241}, \href
  {http://adsabs.harvard.edu/abs/2015Natur.518..512W} {518, 512}

\makeatother
\end{thebibliography}

%%%%%%%%%%%%%%%%%%%%%%%%%%%%%%%%%%%%%%%%%%%%%%%%%%

%%%%%%%%%%%%%%%%% APPENDICES %%%%%%%%%%%%%%%%%%%%%

\appendix

\section{Appendix}
\subsection{Alternate density profiles} 
\label{subsec:alternates}
%\begin{figure}
%\includegraphics[width=\columnwidth]{FINAL2/c_alts.pdf}
%\vspace{2 mm}
%\caption{The escape velocity is not sensitive to the halo concentration ($\lesssim$ 0.1 dex increase for a factor of $\approx$ 2.2 increase in $c$).}
%\label{fig:c_alts}
%\end{figure}
\begin{figure}
\vspace{15 mm}
\includegraphics[width=\columnwidth]{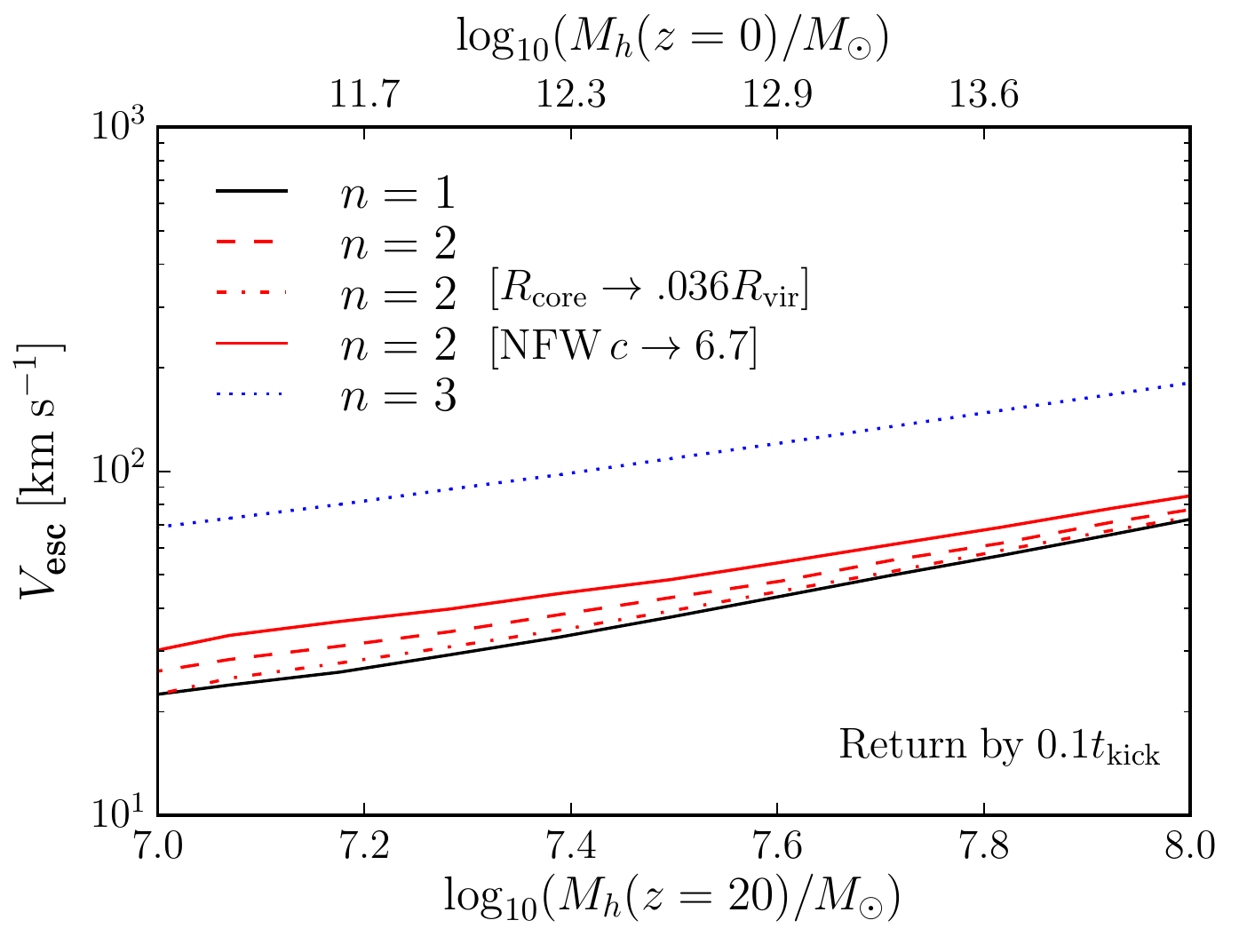}
\vspace{1 mm}
\caption{Effect of variations in the DM and gas density profiles. At $z=20$, the NFW concentration is $c \approx 3$ for all halos. Immediately after a major merger the halo concentration increases to $c \approx 6.7$, leading to $\lesssim 0.1$ dex increase in the velocity needed to escape. Increasing the size of the gas core from 1 pc (fiducial value) to .036$\Rvir$ ($\approx$ 25 pc for $\Mh = 10^{8} \Msun$) has similarly minor effects on $\Vesc$.  However, the escape velocity is sensitive to variations in the
power-law index $n$ of the gas profile $\left( \rho(r) = \widetilde{\rho}_0r^{-n}\right)$ in the range $n=2$ to $n=3$. }
\label{fig:gas_alts}
\end{figure}
\begin{figure}
\includegraphics[width=\columnwidth]{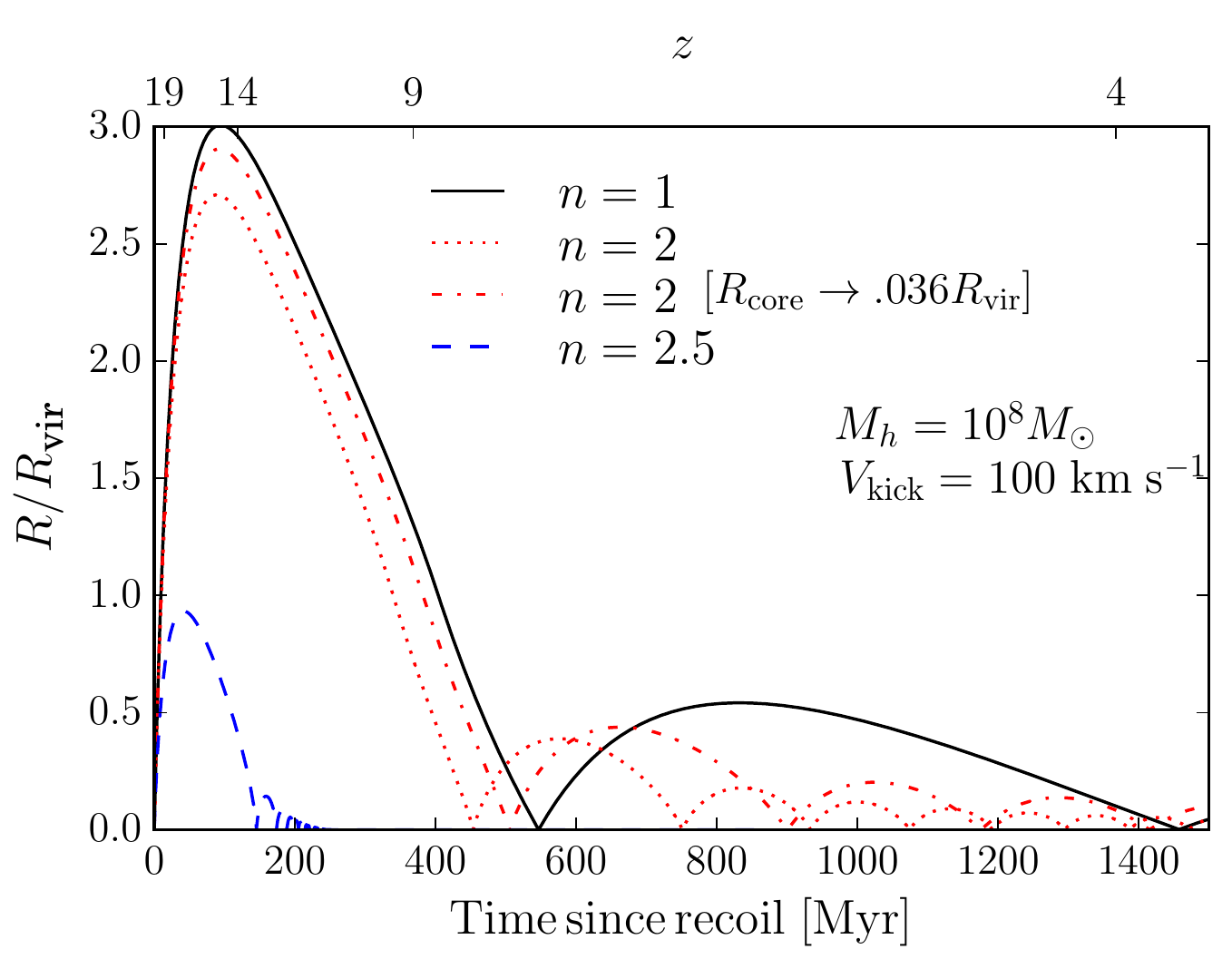}
\vspace{-5 mm}
\caption{SMBH trajectories for different gas profiles. Increasing the size of the gas core has little effect on the shape of the orbit. The largest change occurs for variations in the power-law index $n$ $\left( \rho(r) = \widetilde{\rho}_0r^{-n}\right)$ in the range $n=2$ to $n=3$.  The kick is at $z=20$ with $\Mbh = 10^{5} \Msun$ and $\Mh = 10^{8} \Msun$.}
\label{fig:orbit_gas_alt}
\end{figure}
\begin{figure}
\includegraphics[width=\columnwidth]{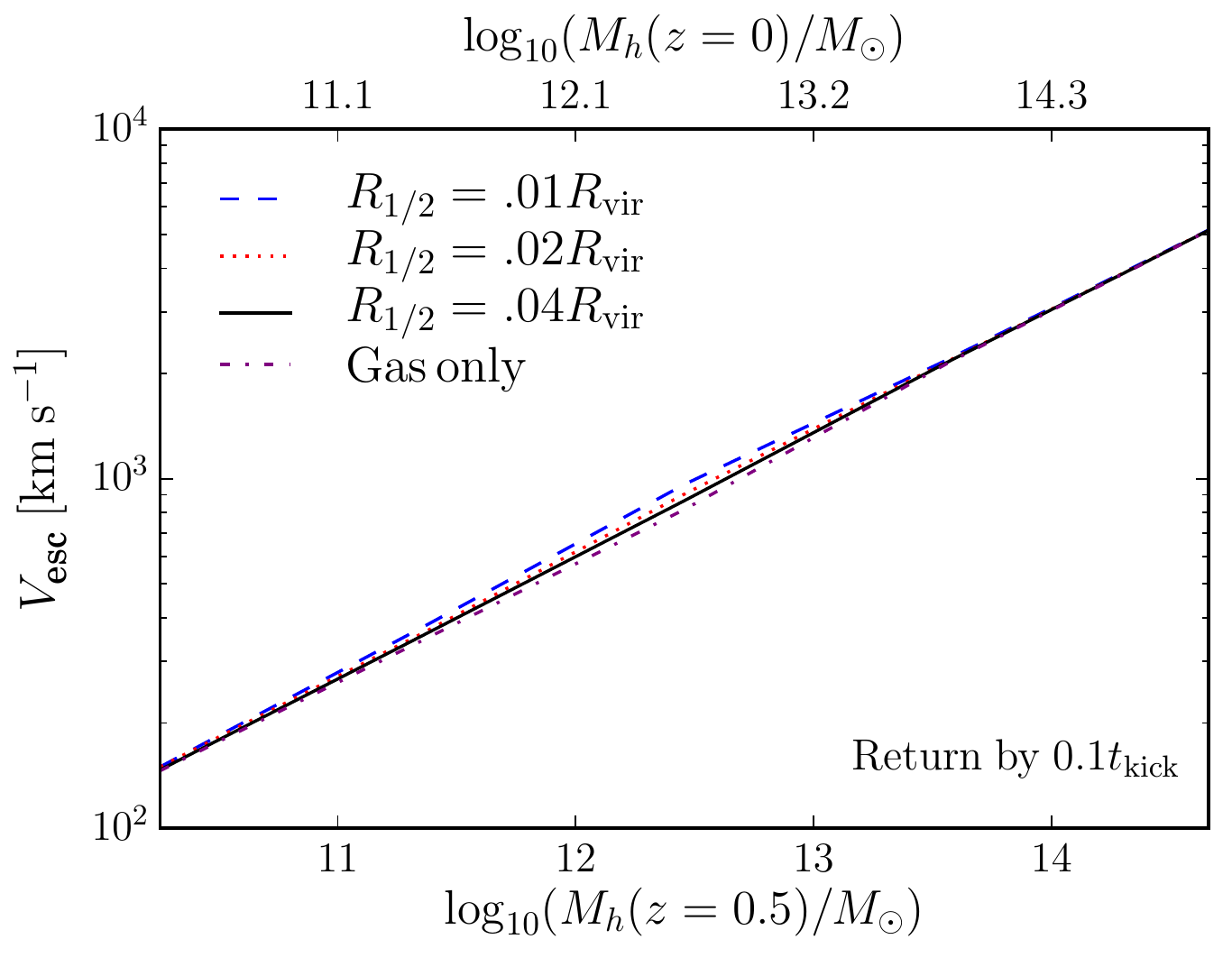}
\vspace{-5 mm}
\caption{Even at $z\sim 0.5$, when the stellar mass is comparable to the gas mass, the velocity needed to escape is not sensitive to stellar distribution.}
\label{fig:stars_alt}
\end{figure}
\begin{figure}
\includegraphics[width=\columnwidth]{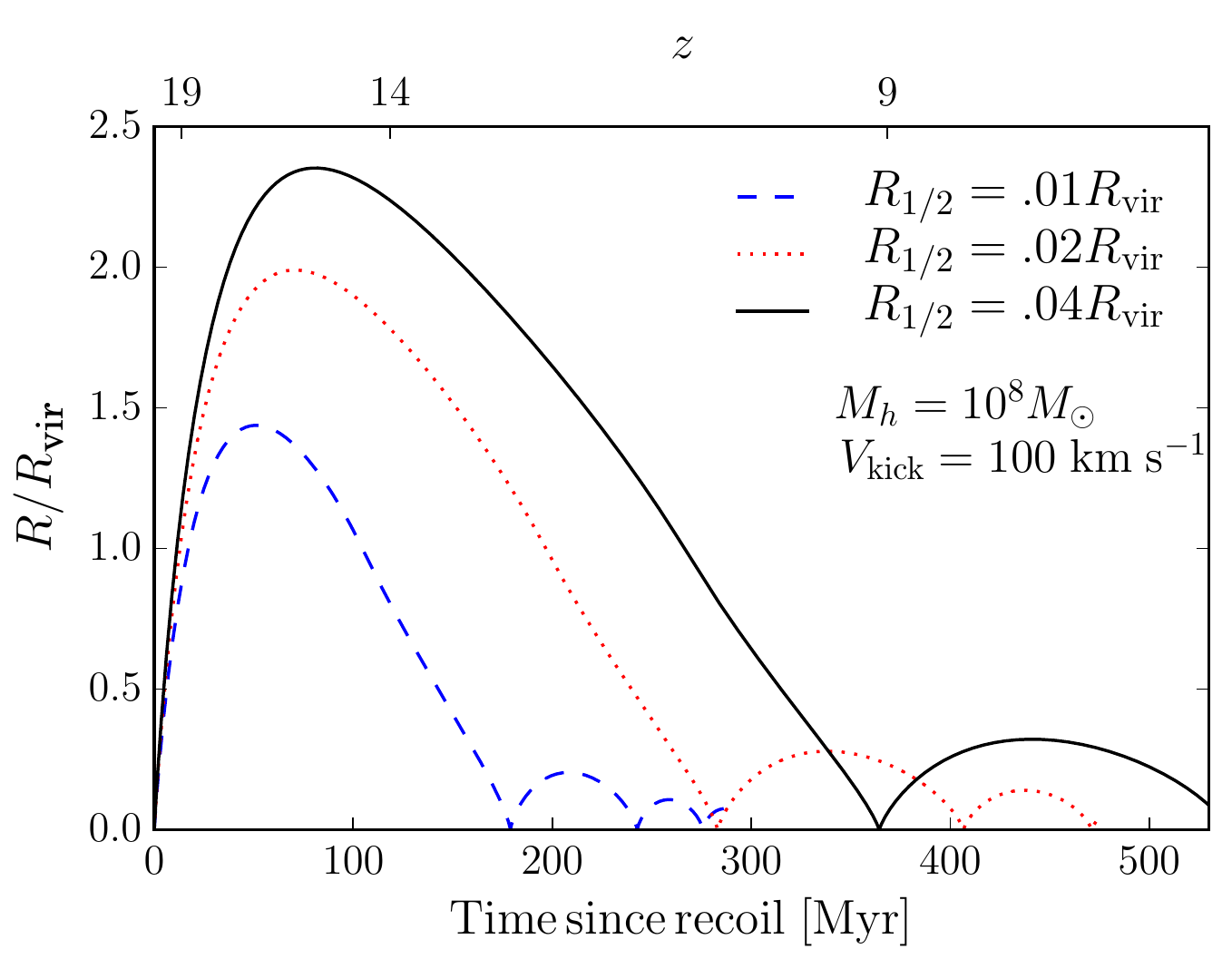}
\vspace{-5 mm}
\caption{SMBH trajectories for different stellar profiles. Smaller half-mass radii deepen the potential, preventing the SMBH from escaping to larger distances. The kick is at $z=20$ with $\Mbh = 10^{5} \Msun$ and $\Mh = 10^{8} \Msun$.}
\label{fig:orbit_stars_alt}
\end{figure}
SMBHs will merge following the mergers of their host halos. Halo mergers will increase the host's central potential for $\approx$1 dynamical time after the merger. A typical 1:3 major merger will lead to an $\sim$10\% increase in the maximum circular velocity ($v_{\mathrm{max}}$) of the halo \citep{behroozi_etal_2014}. The initial and final concentrations are related by:
\begin{equation}
 \frac{c}{\ln(1+c) - \frac{c}{1+c}} = \frac{1}{1.1^2}\times \frac{c'}{\ln(1+c') - \frac{c'}{1+c'}}
 \label{eqn:deltac}
\end{equation}
At $z \sim 20$, the concentration varies only weakly with halo mass and $c \approx 3$ for all halos. Eq. \ref{eqn:deltac} yields $c' \approx 6.7$ after the merger. However, we find only a minor change of $\lesssim$ 0.1 dex increase in escape velocities (Fig. \ref{fig:gas_alts}).

Mergers will also funnel gas to the center of the halo, increasing the central gas density. Alternatively, halos are susceptible to feedback creating a shallower central gas density profile. We therefore test power-law gas profiles, $\rho(r) = \widetilde{\rho}_0r^{-n}$, with $n$ ranging from -1 to -3, as well as different sizes for the central gas cores (Fig. \ref{fig:gas_alts}). We find significantly different behaviour for $n \lesssim 2$ and $n \gtrsim 2$. In both regimes, larger $n$ makes escape to larger radii more difficult and increases the escape velocity (Fig. \ref{fig:orbit_gas_alt}). However, only minor changes in escape velocities and trajectories are observed for $n \lesssim 2$. In contrast, the maximum radial distance achieved decreases rapidly for $n \gtrsim 2$. In this regime, the escape velocity varies by $\lesssim 0.4$ dex. An analytic computation of the escape velocity for both $n=2$ and $n=3$ shows the difference is largely caused by a deeper potential due to a more concentrated gas profile. However, simulations suggest that such steep slopes are rarely achieved. E.g., \cite{moster_etal_2012} find that $n$ increases from $\sim$1.8 to $\sim$1.9 between the pre-merger gas profile and the profile at first coalescence of the two galaxies, thus keeping $\Vesc$ well-approximated by our $n=2.2$ fiducial model.

Similarly, \cite{somerville_etal_2014} find 0.25 dex scatter in the stellar half-mass radius, around the median value of $.02R_{\mathrm{vir}}$. Even at low redshift when more stars are present, including or varying the stellar profile has negligible effect on the escape velocity (Fig. \ref{fig:stars_alt}). However, as with the gas profile, smaller half-mass radii deepen the potential and decrease the maximum radial distance achieved by the SMBH (\ref{fig:orbit_stars_alt}).

%%%%%%%%%%%%%%%%%%%%%%%%%%%%%%%%%%%%%%%%%%%%%%%%%%

% Don't change these lines
\bsp	% typesetting comment
\label{lastpage}
\end{document}